%% file: main.tex
\documentclass[sigconf,screen,nonacm]{acmart}

\AtBeginDocument{%
  }

\usepackage{trfrac}

\usepackage{enumitem}
\usepackage{fancybox}
\usepackage{tikz}

\usepackage{amsmath,amssymb}
\usepackage{balance}

\usepackage{color, colortbl}
\definecolor{Gray}{gray}{0.85}
\definecolor{verylightgray}{rgb}{.97,.97,.97}
\definecolor{mygreen}{RGB}{24,141,31}
\definecolor{myred}{RGB}{142,0,8}
\definecolor{mypurple}{RGB}{107,29,111}

\usepackage{hyperref}
\usepackage{booktabs, threeparttable}
\usepackage[normalem]{ulem}
\useunder{\uline}{\ul}{}
\usepackage{bbding}
\usepackage{multirow}
\usepackage{amsfonts}
\usepackage{listings}
\usepackage{tcolorbox}
\usepackage{caption,subcaption}
\usepackage{comment}
\usepackage[linesnumbered,ruled,noend]{algorithm2e}

\SetKwRepeat{Do}{do}{while}

\usepackage{wasysym}
\usepackage{pifont}
\newcommand*\circlednum[1]{\raisebox{.5pt}{\textcircled{\raisebox{-.9pt} {#1}}}}

\newcommand{\lifter}{\textsc{ActLifter}}
\newcommand{\cluster}{\textsc{ActCluster}}
\newcommand{\hunter}{\textsc{MEVHunter}}

\newcommand{\mevformula}[1]{\textsl{\footnotesize #1}}
\newcommand{\mevsmallformula}[1]{\textsl{\scriptsize #1}}
\newcommand{\mevtinyformula}[1]{\textsl{\tiny #1}}

\newlength\mylen

\newcommand{\ignore}[1]{}

\usepackage{algorithmic}
\usepackage{graphicx}
\usepackage{textcomp}
\usepackage{xcolor}

\input{solidity-highlighting.tex}

\begin{document}
\title{Demystifying DeFi MEV Activities in Flashbots Bundle} %

\author{Zihao Li}
\affiliation{
The Hong Kong Polytechnic University
\country{China}
}

\author{Jianfeng Li}
\affiliation{Xi’an Jiaotong University
\country{China}}

\author{Zheyuan He}
\affiliation{ University of Electronic Science and
Technology of China 
\country{China}}

\author{Xiapu Luo}
\authornote{Corresponding authors}
\affiliation{
The Hong Kong Polytechnic University \country{China}
}

\author{Ting Wang}
\affiliation{ Pennsylvania State University \country{USA}}

\author{Xiaoze Ni}
\affiliation{University of Electronic Science and
Technology of China \country{China}}

\author{Wenwu Yang }
\affiliation{University of Electronic Science and
Technology of China \country{China}
}

\author{Xi Chen}
\affiliation{University of Electronic Science and
Technology of China \country{China}}

\author{Ting Chen\footnotemark[1]{}}
\affiliation{University of Electronic Science and
Technology of China \country{China}
}

\renewcommand{\shortauthors}{Zihao Li et al.}

\begin{abstract}
Decentralized Finance, mushrooming in permissionless blockchains,
has attracted a recent surge in popularity.
Due to the transparency of permissionless blockchains, opportunistic traders can compete to earn revenue
by extracting Miner Extractable Value (MEV), which undermines both the consensus security and efficiency of blockchain systems.
The Flashbots bundle mechanism further aggravates the MEV competition because
it empowers opportunistic traders with the capability of designing more sophisticated MEV extraction.
In this paper, we conduct the first systematic study on DeFi MEV activities in Flashbots bundle by developing \lifter{}, a novel automated tool for accurately identifying DeFi actions in transactions of each bundle, and \cluster{}, a new approach that leverages iterative clustering to facilitate us to discover known/unknown DeFi MEV activities. 
Extensive experimental results show that \lifter{} can achieve nearly 100\% precision and recall in DeFi action identification, significantly outperforming state-of-the-art techniques. Moreover, with the help of \cluster{},  we obtain many new observations and discover 17 new kinds of DeFi MEV activities, which occur in 53.12\% of bundles but have not been reported in existing studies.
\end{abstract}



\maketitle

\section{Introduction}
\label{sec_intro}

Decentralized Finance (DeFi) has attracted a recent surge in popularity
with more than 40B USD total locked value~\cite{defipulse2021defipulse}.
Since transactions broadcasted in the underlying P2P network of blockchain are globally visible,
opportunistic traders can strategically adjust the gas price to prioritize their transactions and earn extra revenue from DeFi, which is known as the generic term Miner Extractable Value (MEV)~\cite{Daian2020flash, Eskandari2020sok, zhou2021high, zhou2021just, ferreira2021frontrunner, qin2021quantifying, wang2021cyclic}.

MEV competition undermines both the security and efficiency of blockchain systems.
First, it incentivizes financially rational validators (miners in the context of PoW) to fork the chain,
thereby deteriorating the blockchain's consensus security~\cite{qin2021quantifying, Daian2020flash}.
Second,
it aggravates network congestion (i.e., P2P network load) and chain congestion (i.e., block space usage) because opportunistic traders who compete for MEV opportunities
prioritize their transactions at the cost of considerable time delay for other transactions~\cite{Daian2020flash}.

The Flashbots organization proposed the bundle mechanism which enables opportunistic traders to design more sophisticated MEV extraction for profits, because it allows traders to submit a sequence of self-constructed and/or selected transactions as a bundle, which can even include unconfirmed transactions broadcasted on the P2P network.
It was reported that
compared to the vanilla Sandwich attacks,
the bundle-based variants were more profitable~\cite{mbaexample2021mbaexample}.

However, little is known about DeFi MEV activities conducted through the bundle mechanism. To demystify the status quo of DeFi MEV activities in bundles, we aim at answering the following questions, namely how prevalent are known DeFi MEV activities in bundles? Are there new DeFi MEV activities that are unreported before in bundles? If that is the case, how did they behave and how prevalent are they? What are the differences between DeFi MEV activities in bundles and other DeFi MEV activities? The answers to these questions can help researchers have an in-depth understanding of DeFi MEV activities, e.g., the features of various MEV activities and the robustness of today's MEV mitigation techniques.

In this paper, we conduct the first systematic study on DeFi MEV activities performed through Flashbots bundle. 
A DeFi MEV activity usually consists of several DeFi actions, each of which refers to an interaction between a trader and an individual function provided by the contracts of DeFi applications. 
For example, a contract of AMM (Automated Market Maker) should support the swap DeFi action for exchanging different assets~\cite{xu2021sok}.
A cyclic arbitrage~\cite{wang2021cyclic} MEV activity involves multiple swap actions in different contracts of AMMs with different prices for profits.

To characterize DeFi MEV activities, we need to first recognize them according to their DeFi actions.
Although existing studies~\cite{wang2021cyclic, qin2021quantifying, qin2021empirical, wang2022speculative, piet2022extracting, wu2021defiranger, explore2021bundles, etherscan2015etherscan,weintraub2022flash} examined DeFi MEV activities and their DeFi actions, they cannot conduct a systematic study on DeFi MEV activities in Flashbots bundle because they suffer from two limitations. 
First, the majority of them~\cite{wang2021cyclic, qin2021quantifying, qin2021empirical, wang2022speculative, piet2022extracting, explore2021bundles,weintraub2022flash} focus on a few DeFi applications and could not be easily extended to cover other DeFi applications because they rely on considerable manual efforts to derive
the rules for
recognizing
DeFi actions according to the specific events emitted by the contracts of DeFi applications and their arguments (cf. Table~\ref{table_related_technique_com}). 
Thus, they will miss many DeFi 
actions
in bundles. 
Although DeFiRanger~\cite{wu2021defiranger} intends to address this limitation by adopting an automated approach to recognize DeFi actions, it suffers from inaccurate recognition of DeFi actions as shown in~\S \ref{sec_rq2}.
Second, none of them can recognize DeFi MEV activities with unknown patterns of DeFi actions.  

\begin{figure}[t!]
	\centering
	\includegraphics[width=0.99\linewidth]{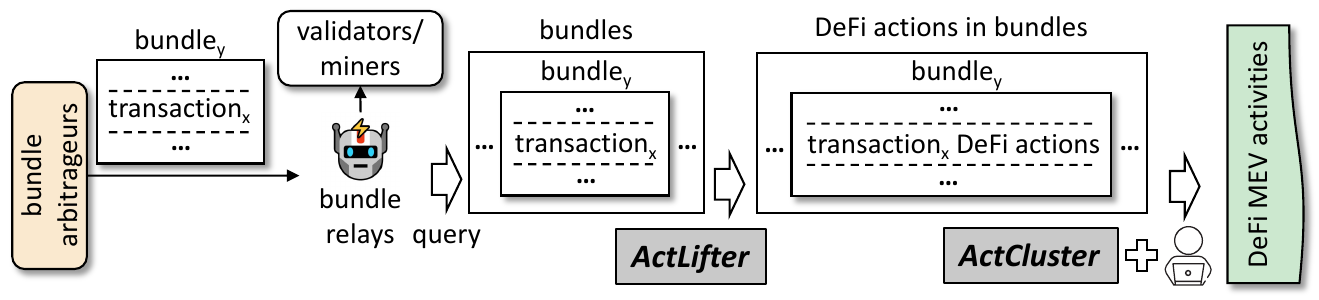}
	\vspace{2pt}
	\caption{Workflow of discovering DeFi MEV activities in bundles. \lifter{} collects the bundles constructed by bundle arbitrageurs and identifies DeFi actions in transactions of each bundle. Inspecting DeFi actions in each bundle, \cluster{} facilitates us to discover DeFi MEV activities.}
    \vspace{2pt}
	\label{fig_discovery}
\end{figure}

To address the aforementioned limitations, we design a new approach shown in  Fig.~\ref{fig_discovery} to discover known and unknown DeFi MEV activities in bundles. 
We first collect bundles constructed by bundle arbitrageurs via querying Flashbots' APIs~\cite{flashbots2021api}.
Then, to address the first limitation, we propose \lifter{}~(\S\ref{sec_actlifter}), a novel automated tool for accurately identifying DeFi actions in the transactions of each bundle.
\lifter{} first recognizes the contracts that operate the DeFi actions, the type of the DeFi actions, and the asset transfers involved in the DeFi actions according to the captured events ~(\S \ref{sec_marketidentification}),
then identifies DeFi actions according to the asset transfer patterns of DeFi actions~(\S \ref{sec_actions_formula}). 
It is worth noting that only a one-off small amount of manual effort is needed to collect events that will be emitted while executing DeFi actions, 
and we provide scripts to automate the process as much as possible (\S \ref{sec_preparation}).   

To address the second limitation, it is inevitable to involve manual inspection to uncover new DeFi MEV activities. To reduce manual efforts, 
we propose \cluster{} (\S \ref{sec_actcluster}), a new approach that uses representation learning~\cite{noroozi2017representation} to 
derive distinguishable feature vectors of bundles according to DeFi actions recognized by \lifter{},
and leverages iterative clustering analysis~\cite{liu2019logzip} and our pruning strategies to facilitate us to discover new DeFi MEV activities.

We conduct extensive experiments (\S \ref{sec_evaluation}) to evaluate the performance of \lifter{} and use \cluster{} to discover DeFi MEV activities from 6,641,481 bundles (from the launch of the bundle mechanism on Feb. 11, 2021 to Dec. 1, 2022).
More precisely, we evaluate the effectiveness of \lifter{} in identifying ten kinds of common DeFi actions and compare it with two state-of-the-art techniques, i.e., Etherscan~\cite{etherscan2015etherscan} and DeFiRanger~\cite{wu2021defiranger}. 
For a fair and convincing comparison with ethical consideration, we spent more than six months in collecting 1,358,122 transactions from Etherscan to mitigate potential risks or negative effects. We queried one page of Etherscan per 10 seconds, which is slower than the human click speed, and manually solved the reCAPTCHA human authentication.
The experimental results show that \lifter{} outperforms existing techniques 
and achieves nearly 100\% precision and recall.
Moreover, we discovered 17 new kinds of DeFi MEV activities and three known DeFi MEV activities in bundles with the help of \cluster{}, which reduced at least 24.2\%, 97.8\% and 98.8\% of manual efforts than three baseline strategies. %

We further demonstrate how our approach (i.e., \lifter{} and \cluster{}) can enhance relays' MEV countermeasures (\S\ref{sec_hunter}), evaluate forking and reorganization (abbr. reorg) risks caused by bundle MEV activities (\S\ref{sec_consensus_security}), and evaluate the impact of bundle MEV activities on blockchain users' economic security (\S\ref{sec_network_security}). Moreover, we discuss three feasible usages of our approach in MEV studies, supported by our experimental results and observations (\S\ref{sec_other_applications}).

\begin{table}[t!]
\small
\centering
\caption{Comparison of \lifter{} and other methods}
\resizebox{0.89\linewidth}{!}{
\begin{threeparttable}
\begin{tabular}{|l|cccccccccc|}
\hline
\multirow{1}{*}{Methods}  %
                                & SW & AL & RL & LI & NM & NB & LE & BO & AI & RE \\ \hline
\rowcolor{black!20} Qin et al.~\cite{qin2021quantifying}               & $\color{mygreen} \checkmark$     & \color{myred} \ding{55}             & \color{myred} \ding{55}               &  $\color{mygreen} \checkmark$ & \color{myred} \ding{55} & \color{myred} \ding{55}  & \color{myred} \ding{55} & \color{myred} \ding{55} & \color{myred} \ding{55} & \color{myred} \ding{55}          \\
\rowcolor{black!5} Qin et al.~\cite{qin2021empirical}                &   \color{myred} \ding{55}   & \color{myred} \ding{55}            & \color{myred} \ding{55}                &  $\color{mygreen} \checkmark$  & \color{myred} \ding{55} & \color{myred} \ding{55}  & \color{myred} \ding{55} & \color{myred} \ding{55} & \color{myred} \ding{55} & \color{myred} \ding{55}          \\
\rowcolor{black!20} Wang et al.~\cite{wang2022speculative} & \color{myred} \ding{55} & \color{myred} \ding{55} & \color{myred} \ding{55} & $\color{mygreen} \checkmark$ & \color{myred} \ding{55} & \color{myred} \ding{55} & $\color{mygreen} \checkmark$ & $\color{mygreen} \checkmark$ & \color{myred} \ding{55} & \color{myred} \ding{55} \\

\rowcolor{black!5} Wang et al.~\cite{wang2021cyclic}           & $\color{mygreen} \checkmark$      & \color{myred} \ding{55}             &  \color{myred} \ding{55}               &  \color{myred} \ding{55}  & \color{myred} \ding{55} & \color{myred} \ding{55}  & \color{myred} \ding{55} & \color{myred} \ding{55} & \color{myred} \ding{55} & \color{myred} \ding{55}          \\
\rowcolor{black!20} Mev-explore~\cite{explore2021bundles}          & $\color{mygreen} \checkmark$     & \color{myred} \ding{55}             &  \color{myred} \ding{55}               & $\color{mygreen} \checkmark$  & \color{myred} \ding{55} & \color{myred} \ding{55}  & \color{myred} \ding{55} & \color{myred} \ding{55} & \color{myred} \ding{55} & \color{myred} \ding{55}          \\
\rowcolor{black!5} Piet et al.~\cite{piet2022extracting}  & $\color{mygreen} \checkmark$ & $\color{mygreen} \checkmark$ & $\color{mygreen} \checkmark$ & \color{myred} \ding{55} & \color{myred} \ding{55} & \color{myred} \ding{55}  & \color{myred} \ding{55} & \color{myred} \ding{55} & \color{myred} \ding{55} & \color{myred} \ding{55}  \\
\rowcolor{black!20} Weintraub et al.~\cite{weintraub2022flash}                   & $\color{mygreen} \checkmark$     & \color{myred} \ding{55}             &  \color{myred} \ding{55}               &  $\color{mygreen} \checkmark$   & \color{myred} \ding{55}  & \color{myred} \ding{55} & \color{myred} \ding{55} & \color{myred} \ding{55} & \color{myred} \ding{55} & \color{myred} \ding{55}       \\
\rowcolor{black!5} Etherscan~\cite{etherscan2015etherscan}                   & $\color{mygreen} \checkmark$     & $\color{mygreen} \checkmark$             &  $\color{mygreen} \checkmark$               &  $\color{mygreen} \checkmark$   & $\color{mygreen} \checkmark$ & $\color{mygreen} \checkmark$  & \color{myred} \ding{55} & $\color{mygreen} \checkmark$ & \color{myred} \ding{55} & \color{myred} \ding{55}        \\
\rowcolor{black!20} DeFiRanger~\cite{wu2021defiranger}           & $\color{mygreen} \checkmark$     & $\color{mygreen} \checkmark$             & $\color{mygreen} \checkmark$                & \color{myred} \ding{55}      & \color{myred} \ding{55} & \color{myred} \ding{55}  & \color{myred} \ding{55} & \color{myred} \ding{55} & \color{myred} \ding{55} & \color{myred} \ding{55}      \\ \hline \hline
\lifter{}                                    &  $\color{mygreen} \checkmark$    & $\color{mygreen} \checkmark$             & $\color{mygreen} \checkmark$                & $\color{mygreen} \checkmark$   & $\color{mygreen} \checkmark$ & $\color{mygreen} \checkmark$  & $\color{mygreen} \checkmark$ & $\color{mygreen} \checkmark$ & $\color{mygreen} \checkmark$ & $\color{mygreen} \checkmark$         \\ \hline
\end{tabular}
\vspace{-1pt}
\begin{tablenotes}[flushleft]
{
\setlength{\itemindent}{-2.49997pt} \small
\item  \footnotesize{\textbf{DeFi actions.} SW: Swap, AL: AddLiquidity, RL: RemoveLiquidity, LI: Liquidation, NM: NFT-Minting, NB: NFT-Burning, LE: Leverage, BO: Borrowing, AI: Airdrop, RE: Rebasing.}
}
\end{tablenotes}
\end{threeparttable}
}
\label{table_related_technique_com}
\end{table}

In summary, this work makes the following contributions.

\begin{itemize}[leftmargin=*,topsep=1pt] %
\item \textit{First systematic analysis of DeFi MEV activities in bundles.} To our best knowledge, 
our work constitutes the first effort toward a systematic analysis of DeFi MEV activities conducted through Flashbots bundle mechanism after tackling two limitations.

\item \textit{Novel approach for identifying DeFi actions.} We propose \lifter{}, a novel approach for automatically identifying DeFi actions from transactions, which outperforms existing techniques and achieves nearly 100\% precision and recall.

\item \textit{New approach for discovering bundle MEV activities.} We propose \cluster{}, a new approach facilitating us to discover bundle MEV activities with much less manual efforts. In particular, it empowers us to discover 17 new kinds of DeFi MEV activities.

\item \textit{New applications.} We demonstrate the usages of our approach (i.e., \lifter{} and \cluster{}), including enhancing relays' MEV countermeasures, evaluating forking and reorg risks caused by bundle MEV activities, and evaluating the impact of bundle MEV activities on blockchain users' economic security. Additionally, we discuss three feasible usages of our approach in MEV studies, supported by our experimental results and observations.

\end{itemize}

\noindent We refer readers to~\cite{appendixpaper} for our full paper version with the appendix.

\section{Background and notation}
\label{sec_background}

This section introduces DeFi applications and actions, Flashbots bundle, and the notations used in this paper. 
For the basic concepts of smart contracts, events, and ERC20/ERC721 standards, we refer readers to other helpful studies~\cite{tikhomirov2017ethereum,macrinici2018smart,heimbach2022sok}.
Besides, the basic concepts of representation learning can be found in Appendix I.

\subsection{DeFi Applications and Actions}
\label{sec_DeFiAppAct}
We focus on ten core DeFi actions of popular DeFi applications (i.e., AMM, Lending, NFT, and Rebase Token) involved in most MEV activities~\cite{ferreira2021frontrunner, qin2021quantifying, wang2021cyclic, wu2021defiranger, qin2021empirical, zhou2021just, zhou2021high, zhou2021a2mm, wang2022speculative}. Each Defi action is represented in the form \mevformula{C$_{\mevsmallformula{DeFi}}$.action$_{\mevsmallformula{type}}$(params)}, where \mevformula{C$_{\mevsmallformula{DeFi}}$}, \mevformula{action$_{\mevsmallformula{type}}$}, and  \mevformula{params} refer to the smart contract implementing the DeFi action, the type and the parameters of the DeFi action, respectively. %

\noindent\textbf{AMM.} It provides functions that allow traders to perform asset exchanges over liquidity pools automatically~\cite{xu2021sok}. Traders can supply or remove their assets with liquidity pools as liquidity providers, and pay a fee to liquidity providers when they exchange assets via an AMM. 
We focus on three core DeFi actions supported by AMMs:
\begin{itemize}[leftmargin=*,topsep=1pt]
    \item \textbf{A1:}
    Swap action \mevformula{AMM.Swap(x$_1$: Asset$_1$, x$_2$: Asset$_2$)}. It performs asset exchange for a trader,
which lets \mevformula{AMM} receive \mevformula{x$_1$} amount of \mevformula{Asset$_1$} and send out \mevformula{x$_2$} amount of \mevformula{Asset$_2$}.

\item \textbf{A2:} AddLiquidity action \mevformula{AMM.AddLiquidity(x$_1$: Asset$_1$, x$_2$: Asset$_2$, ..., x$_n$: Asset$_n$)} (\mevformula{n} $>$ 0). It lets \mevformula{AMM} receive \mevformula{n} assets from a liquidity provider.
\item \textbf{A3:} RemoveLiquidity action \mevformula{AMM.RemoveLiquidity(x$_1$: Asset$_1$, x$_2$: Asset$_2$, ..., x$_n$: Asset$_n$)} (\mevformula{n} $>$ 0). It lets \mevformula{AMM} return \mevformula{n} assets to a liquidity provider.
\end{itemize}

\noindent\textbf{Lending.} It
provides loanable assets through collateralized deposits~\cite{werner2021sok, bartoletti2021sok,aave2021aave,compound2021compound,makerdao2021makerdao}. 
With a collateralized deposit, a borrower can take loanable crypto assets 
from Lendings. 
It uses two kinds of debt mechanisms~\cite{aave2021aave,compound2021compound,makerdao2021makerdao}, i.e., over-collateralization
and under-collateralization, meaning that borrowers can deposit collateral assets with a higher (resp. lower) value than that of borrowed assets.
We focus on three major DeFi actions supported by Lendings:
\begin{itemize}[leftmargin=*,topsep=1pt]
\item \textbf{A4:} Borrowing action \mevformula{Lending.Borrowing(x$_1$: Asset$_1$)}. It lets a borrower loan \mevformula{Asset$_1$} from \mevformula{Lending} with the over-collateral deposit~\cite{wang2022speculative}.
\item \textbf{A5:} Leverage action \mevformula{Lending.Leverage(x$_1$: Asset$_1$)}. It lets a borrower loan \mevformula{Asset$_1$} from \mevformula{Lending} with the under-collateral deposit~\cite{wang2022speculative}.
\item \textbf{A6:} Liquidation action \mevformula{Lending.Liquidation(x$_1$: Asset$_1$, x$_2$: Asset$_2$)}. It lets a trader send the debt \mevformula{Asset$_1$} to \mevformula{Lending} for repaying the debt asset and receive the collateral \mevformula{Asset$_2$} from the \mevformula{Lending} if the negative price fluctuation of the collateral asset happens~\cite{qin2021empirical}.
\end{itemize}

\noindent
\textbf{NFT (Non-Fungible Token).} It provides unique tokens to represent someone’s ownership of specific crypto assets, e.g., CryptoKitties, or a physical asset, like an artwork~\cite{das2021understanding}. 
Most NFT contracts follow the ERC721 standard~\cite{Entriken2015eip721}.
We focus on two major DeFi actions supported by NFT contracts:
\begin{itemize}[leftmargin=*,topsep=1pt]
\item \textbf{A7:} NFT-Minting action \mevformula{C$_{\mevsmallformula{NFT}}$.NFT-Minting(tokenId$_{x_1}$: Asset$_{\mevsmallformula{C}_{\mevtinyformula{NFT}}}$)}. It lets the NFT contract \mevformula{C$_{\mevsmallformula{NFT}}$} mint an NFT  with the tokenId \mevformula{x$_1$}.

\item \textbf{A8:} NFT-Burning action \mevformula{C$_{\mevsmallformula{NFT}}$.NFT-Burning(tokenId$_{x_1}$: Asset$_{\mevsmallformula{C}_{\mevtinyformula{NFT}}}$)}. It lets the NFT contract \mevformula{C$_{\mevsmallformula{NFT}}$} burn an NFT with the tokenId \mevformula{x$_1$}.
\end{itemize}

\noindent
\textbf{Airdrop.}
The airdrop is a promotional activity for bootstrapping a cryptocurrency project by spreading awareness about the cryptocurrency project~\cite{victor2020address}. A small amount of the cryptocurrency is sent to active users for free when they retweet the post sent by the project account.
We focus on the following action:
\begin{itemize}[leftmargin=*,topsep=1pt]
\item \textbf{A9:} Airdrop action \mevformula{C$_{\mevsmallformula{Airdrop}}$.Airdrop(x$_1$ : Asset$_1$)}. It lets the contract \mevformula{C$_{\mevsmallformula{Airdrop}}$} send out the \mevformula{Asset$_1$}. %
\end{itemize}

\noindent
\textbf{Rebase Token.}
Rebase Token follows a continuous rebasing about the number of 
tokens in circulation (e.g., total supply in ERC20 standard)~\cite{schar2021decentralized,Ampleforth2021Ampleforth}.
For example, token holders' balances increase or decrease automatically according to the token’s price evolution provided by price oracles~\cite{Adams2020UniswapVC}. We focus on the following action:
\begin{itemize}[leftmargin=*,topsep=1pt]
\item \textbf{A10:} Rebasing action \mevformula{C$_{\mevsmallformula{Rebasing}}$.Rebasing()}. It lets token holders' balances in \mevformula{C$_{\mevsmallformula{Rebasing}}$} contract automatically increase or decrease. 
\end{itemize}

\subsection{Flashbots bundle}
\label{sec_back_bundle}

The Flashbots~\cite{flashbot2021bundles} designed the bundle mechanism in 2021.
When transactions broadcast over the P2P network, bundle arbitrageurs can observe and analyze them, and include them into bundles along with other transactions. 
Besides, bundle arbitrageurs can adjust the order of transactions in bundles.
Bundle arbitrageurs then send bundles to trusted relays privately, such as relays of Flashbots~\cite{flashbot2021bundles}, Eden~\cite{eden2021bundles}, and BloXroute~\cite{bloxroute2021bundles}.
The relays distribute bundles to connected miners privately.
During the distribution, %
bundles cannot be observed by other P2P peers,
until bundles are included into blocks.
The connected miners will preferentially include the bundles that are the most profitable to them into the head of their mining blocks by calculating a bundle pricing formula~\cite{flashbot2021bundles}.

Ethereum changed its consensus mechanism from PoW to PoS in September 2022~\cite{ethereum2022merge}.
In the context of PoS, validators are selected to create new blocks and add blocks to the Ethereum blockchain, while miners do these tasks in the context of PoW.
After the Merge, the Proposer-Builder Separation (PBS)~\cite{yang2022sok} is introduced on Ethereum.
In PBS, the role of validators is divided into builders and proposers.
Specifically, 
builders create blocks with transactions from their mempool~\cite{tikhomirov2017ethereum} and proposers submit blocks to the blockchain.

Flashbots proposed MEV-Boost~\cite{mevboost2022} in 2022, which supports bundle mechanism in the context of PBS.
In MEV-Boost, bundles are first propagated from arbitrageurs to builders privately.
After creating blocks with bundles, builders submit blocks to relays privately with promising payments to proposers.
Relays then distribute received blocks to connected proposers privately, and proposers finally pick the block with the most payments to submit to the blockchain.
Currently, Flashbots~\cite{flashbot2021bundles}, Eden~\cite{eden2021bundles}, and BloXroute~\cite{bloxroute2021bundles} maintain their builders and relays based on MEV-Boost.
Besides, 68\% of Ethereum blocks are created and relayed by MEV-Boost~\cite{mevboost2022} from the starting date of MEV-Boost (Sep. 2022) to Jan. 2023~\cite{mevboost2022dashboard}, and 77\% of MEV-Boost blocks (i.e., blocks that are created and relayed by MEV-Boost) are from Flashbots~\cite{flashbotsmevboost2022dashboard}.
Our studies shed light on DeFi MEV activities in bundles, since in both the context of PoW and PoS: i) arbitrageurs construct bundles, ii) bundles are relayed from arbitrageurs to validators/miners privately, and iii) validators/miners submit the most profitable bundles to them into the blockchain. 

\begin{figure}[t!]
	\centering
	\includegraphics[width=0.99\linewidth]{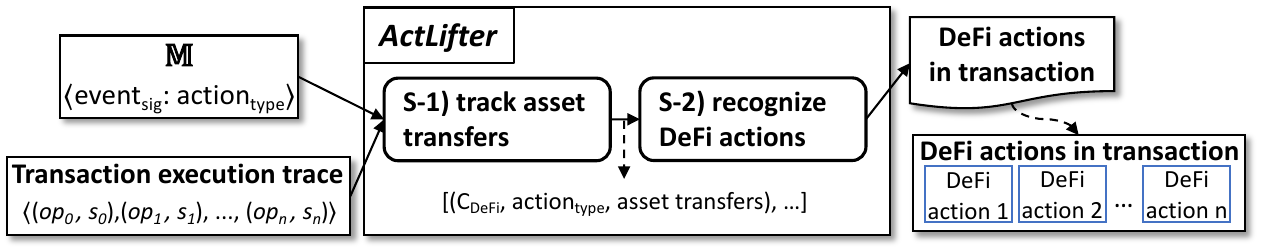}
	\vspace{1pt}
	\caption{Overview of \lifter{}.}
	\label{fig_lifter_overview}
\end{figure}

\subsection{Notation}
\label{sec_notation}

\noindent\textbf{DeFi action.} It, denoted as $A$, represents an interaction between a trader and a function provided by DeFi applications. 
We focus on ten kinds of DeFi actions (\textbf{A1}-\textbf{10} in \S\ref{sec_DeFiAppAct}). %

\noindent\textbf{DeFi actions in a transaction.} A transaction can trigger the execution of multiple contracts, by invoking their functions via internal transactions. Hence, multiple DeFi actions can be operated in a transaction.
We use $\mathbb{A}$ to denote a sequence of \mevformula{n (n $>$ 0)} 
DeFi actions operated in a transaction, where \mevformula{$\mathbb{A}$ = [A$_1$, A$_2$, ..., A$_n$]}. Note that these DeFi actions will be executed one by one in the order. %

\noindent\textbf{DeFi actions in a bundle.} 
A bundle includes a sequence of \mevformula{m (m $>$ 0)} 
transactions, each of which can be signed by different accounts. These transactions will be executed one by one in the order. We use \mevformula{$\mathbb{B}$ = [$\mathbb{A}_1$, $\mathbb{A}_2$, ..., $\mathbb{A}_m$]} to denote all the DeFi actions involved in a bundle.

\noindent\textbf{Asset transfer.}
Since DeFi actions involve one or more asset transfers according to their definitions~\cite{wang2022speculative, bartoletti2021sok, xu2021sok, schar2021decentralized,victor2020address}, to identify DeFi actions, we need to recognize asset transfers and match them against asset transfer patterns of DeFi actions.
We denote an asset transfer as \mevformula{Asset.Transfer(From, To, Value)}, which means \mevformula{From} transfers \mevformula{Value} amount of \mevformula{Asset} to \mevformula{To}, and we consider \mevformula{Asset}, \mevformula{From}, \mevformula{To}, and \mevformula{Value} are parameters of asset transfers. 
We consider two kinds of assets, i.e., crypto token and ETH, and 
distinguish them by the subscript of \mevformula{Asset}. For example, \mevformula{Asset$_{\mevsmallformula{Ether}}$} refers to the ETH asset and \mevformula{Asset$_\mevsmallformula{C}$} refers to the token asset maintained by the contract \mevformula{C}.
We focus on ERC20 and ERC721 token assets, and our approach can be easily extended to other token assets by recognizing asset transfers from their standard events.
We denote ERC721 token minting/burning as \mevformula{Asset}$^{721}_{\mevsmallformula{C}}$\mevformula{.Minting/Burning(From, To, Value)}, meaning that the contract \mevformula{C} mints/burns an NFT of \mevformula{Asset}$^{721}_{\mevsmallformula{C}}$ with the tokenId \mevformula{Value}. With them, we can recognize the NFT-Minting and NFT-Burning actions. %

\noindent\textbf{Execution trace of a transaction.} It refers to a sequence of states and opcodes executed in a transaction, denoted as \mevformula{$\langle$ (op$_0$, s$_0$), (op$_1$, s$_1$), ..., (op$_n$, s$_n$) $\rangle$}. Each state \mevformula{s$_i$} is in the form of \mevformula{$\langle$ Stack$_i$, Memory$_i$ $\rangle$}, where \mevformula{Stack$_i$} and \mevformula{Memory$_i$} are the stack~\cite{wood2014ethereum} variables and memory~\cite{wood2014ethereum} variables, respectively. %
The opcode \mevformula{op$_i$} is defined in~\cite{wood2014ethereum}. 
For each opcode \mevformula{op$_i$}, the state \mevformula{s$_i$} represents the execution environment~\cite{wood2014ethereum} of \mevformula{op$_i$}, and the state \mevformula{s$_{i+1}$} denotes the state after executing \mevformula{op$_i$}.

\section{\lifter{}}
\label{sec_actlifter}
\begin{figure}[t!]
	\centering
	\includegraphics[width=0.99\linewidth]{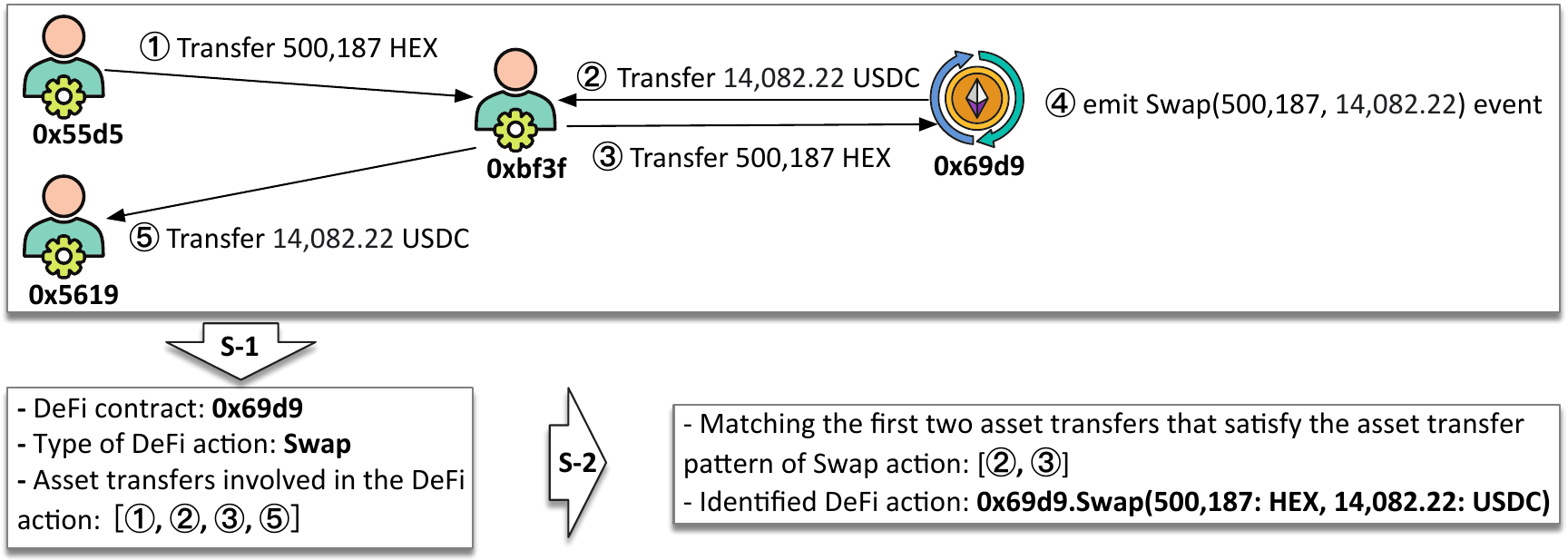}
	\vspace{1pt}
	\caption{\lifter{} identifies a Swap action in a transaction.
	}
	\label{fig_motivating2trace}
\end{figure}

\subsection{Overview}
\label{sec_lifter_overview}

As shown in Fig.~\ref{fig_lifter_overview}, \lifter{} takes in transaction execution trace and the mapping $\mathbb{M}$ between the signature of events and the type of DeFi actions, which is constructed by a semi-automated preparation process (\S \ref{sec_preparation}), and then determines DeFi actions (\textbf{A1-10} in \S \ref{sec_DeFiAppAct}) in the transaction by two steps.

\noindent$\bullet$ \textbf{S-1} (\S \ref{sec_marketidentification}) 
It first locates the emitted events in the execution trace whose signatures are in $\mathbb{M}$.
For each event, it outputs: 
i) the contract that conducts the DeFi action and emits the event,
ii) the corresponding type of the DeFi action in $\mathbb{M}$, %
and iii) the asset transfers involved in the DeFi action. %

\noindent$\bullet$ \textbf{S-2} (\S \ref{sec_actions_formula}) Given the information of each event (i.e., the contract, the type of DeFi action, and the asset transfers), it recognizes the corresponding DeFi action according to the asset transfer patterns (\S \ref{sec_actions_formula}) and outputs them. %

\noindent\textbf{Motivating example.} Fig.~\ref{fig_motivating2trace} shows how \lifter{} identifies a Swap action in a transaction, where there are four asset transfers (i.e., \circlednum{1}, \circlednum{2},
\circlednum{3}, and \circlednum{5}), and the AMM \texttt{\small 0x69d9} emits an event \texttt{\small Swap(}\mevformula{500,187, 14,082.22}\texttt{\small)} in
\circlednum{4}.

In \textbf{S-1}, \lifter{} first locates the event \texttt{\small Swap(}\mevformula{500,187, 14,082.22}\texttt{\small)} emitted in \circlednum{4} whose signature is in $\mathbb{M}$, and then recognizes i) the contract \texttt{\small 0x69d9} since it emits the event in \circlednum{4}, ii) the type of DeFi action (i.e., Swap), and iii) the four asset transfers (i.e., \circlednum{1}, \circlednum{2}, \circlednum{3}, and \circlednum{5}) since their parameters are logged in the event, i.e., \mevformula{500,187} and \mevformula{14082.22}. 
The outputs of \textbf{S-1} are \mevformula{[(0x69d9, }\mevformula{Swap}\mevformula{, [\circlednum{1}, \circlednum{2}, \circlednum{3}, \circlednum{5}])]}.
In \textbf{S-2}, \lifter{} recognizes the Swap action \mevformula{0x69d9.Swap(500,187: HEX, 14082.22: USDC)} in the transaction according to the asset transfer patterns of Swap action (\S \ref{sec_actions_formula}) (i.e., the contract \texttt{\small 0x69d9}, which operates the Swap action, receives \texttt{\small HEX} in \circlednum{3} and sends out \texttt{\small USDC} in \circlednum{2}).

DeFiRanger~\cite{wu2021defiranger} will report a false Swap action by pairing asset transfers in \circlednum{1} and \circlednum{5}, because it only matches the first two asset transfers that satisfy the criteria defined in \cite{wu2021defiranger}, i.e., one account receives and sends out different assets in two asset transfers.

\begin{algorithm}[t!]
\caption{S-1}
\label{alg_preprocessing}
\scriptsize{
  \KwIn{\mevformula{t}, transaction execution trace}
  \KwIn{$\mathbb{M}$, mapping between the event signatures and the type of DeFi actions}
  \KwOut{\mevsmallformula{[(C, action$_{\mevsmallformula{type}}$, assetTransfers)]}, \mevformula{C} operates a DeFi action, \mevsmallformula{action$_{\mevsmallformula{type}}$} is the DeFi action's type, and \mevsmallformula{assetTransfers} are asset transfers involved in the DeFi action}

\mevformula{output} $\leftarrow$ []\\
\mevformula{events} = $\texttt{\scriptsize ParseEvents}$(\mevformula{t})\\
\For{\mevformula{event$_{\mevsmallformula{i}}$} $\in$ \mevformula{events}}{
    \If{\mevformula{event$_{\mevsmallformula{i}_{\mevtinyformula{sig}}}$} $\in$ $\mathbb{M}$}{
        \mevformula{C$_{\mevsmallformula{i}}$} = $\texttt{\scriptsize GetContract}$(\mevformula{event$_{\mevsmallformula{i}}, \mevformula{t}$})\\
        \mevformula{action$_{\mevsmallformula{type}_{\mevtinyformula{i}}}$} = $\texttt{\scriptsize GetActionType}$(\mevformula{event$_{\mevsmallformula{i}_{\mevtinyformula{sig}}}$}, $\mathbb{M}$)\\
        \mevformula{assetTransfers$_{\mevsmallformula{i}}$} = $\texttt{\scriptsize GetAssetTransfers}$(\mevformula{event$_{\mevsmallformula{i}}$, t})\\
        \mevformula{output}.append((\mevformula{C$_\mevsmallformula{i}$, action$_{\mevsmallformula{type}_{\mevtinyformula{i}}}$, assetTransfers$_{\mevsmallformula{i}}$}))\\
    }
}
\KwRet \mevformula{output}\\

\SetFuncSty{textbf}
\SetArgSty{text} 
\SetKwFunction{GetAssetTransfers}{\texttt{\scriptsize GetAssetTransfers}}
  \SetKwProg{Fn}{Function}{:}{}
  \Fn{\GetAssetTransfers{\mevformula{event$_{\mevsmallformula{i}}$, t}}}{
        \mevformula{retAssetTransfers} $\leftarrow$ []\\
        \mevformula{assetTransfersinTrace} = $\texttt{\scriptsize GetAllAssetTransfers}$(\mevformula{t}) \\
        \For{\mevformula{assetTransfer$_\mevsmallformula{j}$} $\in$ \mevformula{assetTransfersinTrace}}{
            \If{\texttt{\scriptsize IsLogged}(\mevformula{assetTransfer$_\mevsmallformula{j}$},\mevformula{event$_\mevsmallformula{i}$}, \mevformula{t})}{  
                \mevformula{retAssetTransfers}.append(\mevformula{assetTransfer$_\mevsmallformula{j}$})\\
            }
        }
        \KwRet \mevformula{retAssetTransfers}
  }
}
\end{algorithm}

\subsection{Preparation}
\label{sec_preparation}

The majority of existing studies~\cite{wang2021cyclic, qin2021quantifying, qin2021empirical, wang2022speculative, piet2022extracting, explore2021bundles} use specific events to recognize DeFi actions, because smart contracts use events to notify others (e.g., users, third-party tools) about their execution (e.g., state changes)~\cite{zhang2021dharcher}.
Motivated by these studies, we construct a mapping $\mathbb{M}$ from the events to the corresponding type of DeFi actions by leveraging the event information from developers, which is scattered in different places, such as each DeFi’s official website, document, or source codes.
We first develop a tool to collect the descriptions of events or the code snippets and comments of events from the websites of popular DeFi applications listed in DeFiPulse~\cite{defipulse2021new} and
Dapp.com~\cite{dapp2021dapp}.
Then, we manually confirm the results to construct the mapping $\mathbb{M}$.

More precisely, if a DeFi application provides documents, we summarize the document template to extract 
the descriptions of events in its documents.
Our tool also checks whether the extracted event indeed exists in the source codes of the DeFi.
If a DeFi application does not provide documents, our tool inspects its source code to extract code snippets that define events (i.e., the keyword \texttt{\small event} representing the start of an event definition, the event's name, and the definition of the event's parameters) in Solidity or Vyper, the comments of events, and functions that emit events.

Two authors read the information of extracted events independently to determine whether the events correspond to DeFi actions
(Appendix A uses two examples to illustrate how we determine the results.). %
After analyzing the collected information, they discuss and adjust results with the help of a third author to resolve conflicts for the sake of minimizing the impacts of human subjectivity.

The whole procedure of manual analysis cost around 18 hours.
We collect 32 and 56 events from the descriptions of events and the code snippets, and the comments of events, respectively.
Specifically, 
we collect 37, 9, 12, 8, 3, 8, 7, and 4 different events for Swap, AddLiquidity, RemoveLiquidity, Liquidation, Leverage, Borrowing, Airdrop, and Rebasing actions, respectively.
Besides, we leverage the standard \texttt{\small Transfer} event in ERC20~\cite{fabian2015eip20} to recognize NFT-Minting and NFT-Burning actions, since the widely used contract templates for NFT (e.g., OpenZeppelin~\cite{OpenZeppelinERC20721template} and chiru-labs~\cite{chirulabsERC721template}) emit the \texttt{\small Transfer} event during NFT minting and burning.

We further investigate the events in $\mathbb{M}$ to estimate how much manual work we reduce compared to existing studies~\cite{wang2021cyclic, qin2021quantifying, qin2021empirical, wang2022speculative, piet2022extracting, explore2021bundles}.
To our best knowledge, previous studies conduct three steps to derive rules for recognizing DeFi actions: i) find out specific events that correspond to DeFi actions. ii) summarize how to recognize DeFi actions from the arguments of the events. iii) find out extra information (e.g., other events or storage variables) to assist in recognizing DeFi actions if they fail in the ii) step.
For the 88 events in $\mathbb{M}$, we find that 
41 events can be used to 
recognize
DeFi actions according to the ii) step, and 47 events need extra manual work at the iii) step.
Compared to the previous studies, we only need to find out the specific events that correspond to DeFi actions, and obviate the need of the manual efforts for the last two steps.

\begin{table}[t!]
\small
\caption{Conditions of asset transfers}
\vspace{1pt}
\centering
\resizebox{\linewidth}{!}{
\begin{tabular}{|l|c|}
\hline
Asset transfer type            & Conditions \\ \hline 
\rowcolor{black!20} Ether transfer  & $\trfrac[]
    {\begin{trgather}
    \mevformula{c}_1\mevformula{: From.CALL(To, Value) } || \mevformula{ TX(From, To, Value)} \\[1pt]
    \mevformula{c}_2\mevformula{: (Value} \neq \mevformula{0)} \land \mevformula{(From} \neq \mevformula{To)}
    \end{trgather}}
    {\begin{trgather}
    \mevformula{Asset}_{\mevsmallformula{Ether}}\mevformula{.Transfer(From, To, Value)}
    \end{trgather}}$ \\ %
\rowcolor{black!5}Token transfer  & $\trfrac[]
    {\begin{trgather}
    \mevformula{c}_1\mevformula{: C.Event(Transfer(From,To,Value))} \\[1pt]
    \mevformula{c}_2\mevformula{: From} \not\in \mevformula{(0x00...00, C)} \land \mevformula{To} \not\in \mevformula{(0x00...00, C)} \land \mevformula{(Value} \neq \mevformula{0)} \land \mevformula{(From} \neq \mevformula{To)}
    \end{trgather}}
    {\begin{trgather}
    \mevformula{Asset}_{\mevsmallformula{C}}\mevformula{.Transfer(From, To, Value)}
    \end{trgather}}$ \\ %
\rowcolor{black!20} ERC721 token minting     & $\trfrac[]{
\begin{trgather}
    \mevformula{c}_1\mevformula{: C.Event(Transfer(From,To,Value))} \\[1pt]
    \mevformula{c}_2\mevformula{: From} \in \mevformula{(0x00...00, C)} \land \mevformula{To} \not\in \mevformula{(0x00...00, C)} \land \mevformula{(C} \models \mevformula{ERC721 standard}) 
    \end{trgather}
    }{\mevformula{Asset}^{721}_{\mevsmallformula{C}}\mevformula{.Minting(From, To, Value)}}$ \\ %

\rowcolor{black!5}ERC721 token burning     & $\trfrac[]{
\begin{trgather}
    \mevformula{c}_1\mevformula{: C.Event(Transfer(From,To,Value))} \\[1pt]
    \mevformula{c}_2\mevformula{: From} \not\in \mevformula{(0x00...00, C)} \land \mevformula{To} \in \mevformula{(0x00...00, C)} \land \mevformula{(C} \models \mevformula{ERC721 standard})
    \end{trgather}}{\mevformula{Asset}^{721}_{\mevsmallformula{C}}\mevformula{.Burning(From, To, Value)}}$ \\ \hline
\end{tabular}
}
\label{table_assettransferrules}
\vspace{1pt}
\end{table}

\subsection{Step \textbf{S-1}}
\label{sec_marketidentification}

Algorithm~\ref{alg_preprocessing} presents
the process of step \textbf{S-1}. Taking in transaction execution trace and $\mathbb{M}$, \lifter{} 
first locates the emitted events whose signatures are in $\mathbb{M}$. Then, for each event, \lifter{} identifies and outputs the information of the corresponding DeFi action, including \mevformula{C$_{\mevsmallformula{DeFi}}$}, \mevformula{action}$_{\mevsmallformula{type}}$, and \mevformula{params} (\textbf{A1-10} in~\S \ref{sec_DeFiAppAct}).

More precisely, \lifter{} locates all emitted events in the trace (Line 2) by 
retrieving the signature and parameters of events from
the execution state (i.e., \mevformula{Stack} and \mevformula{Memory}) of the opcodes used to log events, i.e., \texttt{\small LOG0-4}~\cite{ethtx}, and only keeps the events whose signatures are in $\mathbb{M}$ (Line 3 and 4). 
\lifter{} also records the contracts that log these events in the trace~\cite{ethtx,chen2019tokenscope,li2023blockexplorer,chen2019dataether,he2023tokenaware} (Line 5) and obtains the type of the corresponding DeFi actions from $\mathbb{M}$ (Line 6).

Since \mevformula{param} of a DeFi action is summarized from asset transfers involved in the DeFi action, \lifter{} tracks asset transfers that are logged by the events through the function \texttt{\small GetAssetTransfers} (Line 7). These asset transfers will be used to recognize DeFi actions in \textbf{S-2}.
We focus on recognizing four kinds of asset transfers described in~\S \ref{sec_notation}, namely Ether transfer, token transfer, and ERC721 token minting/burning. %
Specifically, Ether can be transferred in two ways: i) the sender is a smart contract and executes the \texttt{\small CALL} opcode~\cite{wood2014ethereum} by setting the recipient and the amount of transferred Ether as its parameters in the \mevformula{stack}, ii) the sender is an externally-owned account (EOA)~\cite{dannen2017introducing} and signs a transaction with setting the recipient and the amount of transferred Ether as its parameters.
Moreover,
if a token transfer, or an ERC721 token minting/burning occurs,
an ERC20 standard \texttt{\small Transfer} event~\cite{fabian2015eip20} will be emitted with the parameters of the sender, the recipient, and the amount of transferred token or the tokenId of minted/burnt ERC721 token, according to the specification of ERC20 standard~\cite{fabian2015eip20} and the widely used contract templates for ERC721 (e.g., OpenZeppelin~\cite{OpenZeppelinERC20721template} and chiru-labs~\cite{chirulabsERC721template}).

\begin{table}[t!]
\caption{Asset transfer patterns of ten DeFi actions}
\small
\vspace{1pt}
\centering
\resizebox{\linewidth}{!}{
\begin{tabular}{|l|c|}
\hline
\begin{tabular}[c]{@{}l@{}}DeFi action type\end{tabular} & Asset transfer patterns \\ \hline
\rowcolor{black!20}Swap            & $\trfrac[]
    {\begin{trgather}
    \mevformula{Asset}_1\mevformula{.Transfer(}\_\mevformula{,}\mevformula{C}_{\mevsmallformula{DeFi}}\mevformula{,x}_1\mevformula{)} \land \mevformula{Asset}_2\mevformula{.Transfer(C}_{\mevsmallformula{DeFi}}\mevformula{,}\_\mevformula{,x}_2\mevformula{)}
    \end{trgather}}
    {\begin{trgather}
    \mevformula{C}_{\mevsmallformula{DeFi}}\mevformula{.Swap(x}_1\mevformula{:Asset}_1\textsf{\footnotesize, x}_2\mevformula{:Asset}_2\mevformula{)}
    \end{trgather}}$ \\ %
\rowcolor{black!5}AddLiquidity    & $\trfrac[]
    {\begin{trgather}
    \mevformula{Asset}_1 \mevformula{.Transfer(}\_ \mevformula{,C}_{\mevsmallformula{DeFi}} \mevformula{,x}_1 \mevformula{)} \land \mevformula{Asset}_2 \mevformula{.Transfer(}\_ \mevformula{,C}_{\mevsmallformula{DeFi}} \mevformula{,x}_2 \mevformula{)} \land  \mevformula{...} \land  \mevformula{Asset}_n \mevformula{.Transfer(}\_ \mevformula{,C}_{\mevsmallformula{DeFi}} \mevformula{,x}_n \mevformula{)}\\
    \end{trgather}}
    {\begin{trgather}
    \mevformula{C}_{\mevsmallformula{DeFi}} \mevformula{.AddLiquidity(x}_1 \mevformula{:Asset}_1 \mevformula{, x}_2 \mevformula{:Asset}_2 \mevformula{, ..., x}_n \mevformula{:Asset}_n \mevformula{)}
    \end{trgather}}$ \\ %
\rowcolor{black!20}RemoveLiquidity & $\trfrac[]
    {\begin{trgather}
    \mevformula{Asset}_1\mevformula{.Transfer(C}_{\mevsmallformula{DeFi}}\mevformula{,}\_\mevformula{,x}_1\mevformula{)} \land \mevformula{Asset}_2\mevformula{.Transfer(C}_{\mevsmallformula{DeFi}}\mevformula{,}\_\mevformula{,x}_2\mevformula{)} \land \mevformula{...} \land  \mevformula{Asset}_n\mevformula{.Transfer(C}_{\mevsmallformula{DeFi}}\mevformula{,}\_\mevformula{,x}_n\mevformula{)}
    \end{trgather}}
    {\begin{trgather}
    \mevformula{C}_{\mevsmallformula{DeFi}}\mevformula{.RemoveLiquidity(x}_1\mevformula{:Asset}_1\mevformula{, x}_2\mevformula{:Asset}_2\mevformula{, ..., x}_n\mevformula{:Asset}_n\mevformula{)}
    \end{trgather}}$  \\ %
\rowcolor{black!5}Leverage    & $\trfrac[]{
\begin{trgather} 
\mevformula{Asset}_1\mevformula{.Transfer(C}_{\mevsmallformula{DeFi}}\mevformula{,}\_\mevformula{,x}_1\mevformula{)}
\end{trgather}
}{\mevformula{C}_{\mevsmallformula{DeFi}}\mevformula{.Leverage(x}_1\mevformula{: Asset}_1\mevformula{)}}$ \\ %
\rowcolor{black!20}Borrowing    & $\trfrac[]{
\begin{trgather} 
\mevformula{Asset}_1\mevformula{.Transfer(C}_{\mevsmallformula{DeFi}}\mevformula{,}\_\mevformula{,x}_1\mevformula{)}
\end{trgather}
}{\mevformula{C}_{\mevsmallformula{DeFi}}\mevformula{.Borrowing(x}_1\mevformula{: Asset}_1\mevformula{)}}$ \\ %
\rowcolor{black!5}Liquidation     & $\trfrac[]
    {\begin{trgather}
    \mevformula{Asset}_1\mevformula{.Transfer(}\_\mevformula{, C}_{\mevsmallformula{DeFi}}\mevformula{,x}_1\mevformula{)} \land  \mevformula{Asset}_2\mevformula{.Transfer(C}_{\mevsmallformula{DeFi}}\mevformula{,} \_\mevformula{, x}_2\mevformula{)}
    \end{trgather}}
    {\begin{trgather}
    \mevformula{C}_{\mevsmallformula{DeFi}}\mevformula{.Liquidation(x}_1\mevformula{:Asset}_1\mevformula{, x}_2\mevformula{:Asset}_2\mevformula{)}
    \end{trgather}}$ \\ %
\rowcolor{black!20}NFT-Minting    & $\trfrac[]{
\begin{trgather} 
\mevformula{Asset}^{721}_{C_{\mevsmallformula{DeFi}}}\textsf{\footnotesize.Minting(\_, }\_\mevformula{, x}_1\textsf{\footnotesize)}
\end{trgather}
}{\mevformula{C}_{\mevsmallformula{DeFi}}\mevformula{.NFT-Minting(tokenId}_{x_1}\mevformula{: Asset}_{C_{\mevsmallformula{DeFi}}}\mevformula{)}}$ \\ %
\rowcolor{black!5}NFT-Burning &    $\trfrac[]{
\begin{trgather} 
\mevformula{Asset}^{721}_{C_{\mevsmallformula{DeFi}}}\textsf{\footnotesize.Burning(}\_\textsf{\footnotesize, \_ , x}_1\textsf{\footnotesize)}
\end{trgather}
}{\mevformula{C}_{\mevsmallformula{DeFi}}\mevformula{.NFT-Burning(tokenId}_{x_1}\mevformula{: Asset}_{C_{\mevsmallformula{DeFi}}}\mevformula{)}}$ \\ %
\rowcolor{black!20}Airdrop    & $\trfrac[]{
\begin{trgather} 
\mevformula{Asset}_1\mevformula{.Transfer(C}_{\mevsmallformula{DeFi}}\mevformula{,}\_\mevformula{,x}_1\mevformula{)}
\end{trgather}
}{\mevformula{C}_{\mevsmallformula{DeFi}}\mevformula{.Airdrop(x}_1\mevformula{: Asset}_1\mevformula{)}}$ \\ %
\rowcolor{black!5}Rebasing    & $\trfrac[]{
\begin{trgather} 
\_ 
\end{trgather}
}{\mevformula{C}_{\mevsmallformula{DeFi}}\mevformula{.Rebasing()}}$ \\ \hline
\end{tabular}
}
\label{table_actsemantics}
\end{table}

We summarize two conditions (i.e., \mevformula{c$_1$} and \mevformula{c$_2$}) for identifying each kind of asset transfer. \mevformula{c$_1$}
checks whether an asset transfer occurs, e.g., a sender transfers Ether to a recipient, or a \texttt{\small Transfer} event is emitted. 
However, 
asset transfers, which do not trigger the actual transfer of assets between the sender and the recipient, can pass the check of \mevformula{c$_1$} (e.g., the transferred amount of asset is zero). Thus, we use
\mevformula{c$_2$} to filter out such asset transfers. Table~\ref{table_assettransferrules} lists the four types of asset transfers and their \mevformula{c$_1$} and \mevformula{c$_2$}, which are elaborated as follows.
Due to the page limit, we describe how we recognize Ether transfers as follows, and introduce the rest in Appendix B.

\noindent$\bullet$
\textbf{Ether transfer.} 
In an Ether transfer \mevformula{Asset}$_{\mevsmallformula{Ether}}$\mevformula{.Transfer(From, To, Value)}, \mevformula{From} sends \mevformula{Value} amount of ETH to \mevformula{To}.
Hence, \mevformula{c$_1$} checks whether an Ether transfer occurs in any of the two cases: i) \mevformula{From} is a contract and executes a \texttt{\small CALL} to transfer \mevformula{Value} amount of Ether to \mevformula{To} (i.e., \mevformula{From.CALL(To, Value)}), ii) \mevformula{From} is an EOA account and signs a transaction to send \mevformula{Value} amount of Ether to \mevformula{To} (i.e., \mevformula{TX(From, To, Value)}).
An Ether transfer should also satisfy both requirements in \mevformula{c$_2$}: i)  \mevformula{From} and \mevformula{To} are different accounts (i.e., \mevformula{From} $\neq$ \mevformula{To}), and ii) the transferred amount \mevformula{Value} is non-zero (i.e., \mevformula{Value} $\neq$ \mevformula{0}). Note that there is no actual transfer of Ether between \mevformula{From} and \mevformula{To} if any requirement is violated.

For all asset transfers identified from the trace, we check whether they are logged by events (Line 12-14). %
Specifically,
we check whether the event's parameters contain the asset transfer's parameters, since an event takes parameters of an asset transfer as its parameters to log the asset transfer.
Asset transfers include four parameters, i.e., \mevformula{Asset}, \mevformula{From}, \mevformula{To}, and \mevformula{Value}.
To check the first three parameters which are of the address type~\cite{solidity2022event,chen2021sigrec,zhao2023deep}, we determine whether there are parameters of address type in the event, and values of the parameters are the same as the first three parameters of the asset transfer.
The \mevformula{Value} parameter denotes the amount of transferred asset, and its type is a 256-bit unsigned integer \cite{fabian2015eip20,wood2014ethereum}.
Since an event can convert \mevformula{Value} to another type (e.g., signed integers~\cite{solidity2022event}) and use the converted one as its parameter, 
we also check 
whether absolute values of parameters in the event are the same as \mevformula{Value}'s value.

\begin{algorithm}[t!]
\caption{S-2}
\label{alg_recognizedefiaction}
\scriptsize{
  \KwIn{\mevsmallformula{[(C, action$_{\mevsmallformula{type}}$, assetTransfers)]}, \mevformula{C} operates a DeFi action, \mevsmallformula{action$_{\mevsmallformula{type}}$} is the DeFi action's type, and \mevsmallformula{assetTransfers} are asset transfers involved in the DeFi action}
  \KwOut{$\mathbb{A}$, DeFi actions in a transaction}

\SetArgSty{text} 
$\mathbb{A}$ $\leftarrow$ []\\
\For{(\mevformula{C$_{\mevsmallformula{i}}$}, \mevformula{action$_{\mevsmallformula{type}_{\mevtinyformula{i}}}$}, \mevformula{assetTransfers$_{\mevsmallformula{i}}$}) $\in$ \mevformula{[(C, \mevformula{action$_{\mevsmallformula{type}}$}, \mevformula{assetTransfers})]}}{
        \mevformula{A} $\leftarrow$ null \\
      \uIf{\mevformula{action$_{\mevsmallformula{type}_{\mevtinyformula{i}}}$} $\in$ ("AddLiquidity", "RemoveLiquidity") } {
        \mevformula{A} = $\texttt{\scriptsize GetActionOfNAssetTransfers}$(\mevformula{C$_{\mevsmallformula{i}}$}, \mevformula{action$_{\mevsmallformula{type}_{\mevtinyformula{i}}}$}, \mevformula{assetTransfers$_{\mevsmallformula{i}}$}) \\
      }
      \uElseIf{\mevformula{action$_{\mevsmallformula{type}_{\mevtinyformula{i}}}$} $\in$ ("Swap", "Liquidation") }{
        \mevformula{A} = $\texttt{\scriptsize GetActionOfTwoAssetTransfers}$(\mevformula{C$_{\mevsmallformula{i}}$}, \mevformula{action$_{\mevsmallformula{type}_{\mevtinyformula{i}}}$}, \mevformula{assetTransfers$_{\mevsmallformula{i}}$}) \\
      }
      \uElseIf{\mevformula{action$_{\mevsmallformula{type}_{\mevtinyformula{i}}}$} $\in$ ("Rebasing") }{
        \mevformula{A} = $\texttt{\scriptsize GetActionOfNullAssetTransfer}$(\mevformula{C$_{\mevsmallformula{i}}$}, \mevformula{action$_{\mevsmallformula{type}_{\mevtinyformula{i}}}$}, \mevformula{assetTransfers$_{\mevsmallformula{i}}$}) \\
      }
      \Else{
        \mevformula{A} = $\texttt{\scriptsize GetActionOfOneAssetTransfer}$(\mevformula{C$_{\mevsmallformula{i}}$}, \mevformula{action$_{\mevsmallformula{type}_{\mevtinyformula{i}}}$}, \mevformula{assetTransfers$_{\mevsmallformula{i}}$}) \\
      }
      
      \If{\mevformula{A} $\neq$ null}{
        $\mathbb{A}$.append(\mevformula{A})\\
      }
    }
\KwRet $\mathbb{A}$\\
}
\end{algorithm}

\subsection{Step \textbf{S-2}}
\label{sec_actions_formula}

Given the information (i.e., the contract that executes a DeFi action, the DeFi action's type, and asset transfers involved in the DeFi action) collected in \textbf{S-1}, \lifter{} determines DeFi actions according to their asset transfer patterns in \textbf{S-2}.
Table~\ref{table_actsemantics} summarizes asset transfer patterns of ten DeFi actions according to their definitions~\cite{wang2022speculative, bartoletti2021sok, xu2021sok, schar2021decentralized,victor2020address}. 
We explain them and describe how \lifter{} leverages patterns to recognize DeFi actions as follows.

\noindent$\bullet$
\textbf{Swap.} 
It involves two asset transfers in the transaction, 
where the \mevformula{C$_{\mevsmallformula{DeFi}}$} receives \mevformula{x$_1$} amount of asset \mevformula{Asset$_1$}, and sends out \mevformula{x$_2$} amount of another asset \mevformula{Asset$_2$}.

\noindent$\bullet$
\textbf{AddLiquidity/RemoveLiquidity.} 
It involves \mevformula{n} asset transfers in the transaction. For each asset transfer, \mevformula{C$_{\mevsmallformula{DeFi}}$} receives (resp. sends out) a different kind of asset \mevformula{Asset$_i$}, whose amount is \mevformula{x$_i$}.

\noindent
$\bullet$
\textbf{Leverage/Borrowing.}
It involves one asset transfer, where \mevformula{C$_{\mevsmallformula{DeFi}}$} sends out \mevformula{x$_1$} amount of \mevformula{Asset$_1$}.

\noindent$\bullet$
\textbf{Liquidation.} 
It involves two asset transfers in the transaction, where the \mevformula{C$_{\mevsmallformula{DeFi}}$} receives \mevformula{x$_1$} amount of \mevformula{Asset$_1$}, and sends out \mevformula{x$_2$} amount of a different asset \mevformula{Asset$_2$}.

\noindent
$\bullet$
\textbf{NFT-Minting/NFT-Burning.}
It involves an ERC721 token minting (resp. burning), where \mevformula{C}$_{\mevsmallformula{DeFi}}$ mints (resp. burns) an NFT with the tokenId \mevformula{x$_1$}.

\noindent
$\bullet$
\textbf{Airdrop.}
It involves one asset transfer, where \mevformula{C$_{\mevsmallformula{DeFi}}$} sends out \mevformula{x$_1$} amount of \mevformula{Asset$_1$}.

\noindent
$\bullet$
\textbf{Rebasing.}
Since no asset transfer is involved in the Rebasing action, for the contract \mevformula{C$_{\mevsmallformula{DeFi}}$} that conducts the Rebasing action, we check whether \mevformula{C$_{\mevsmallformula{DeFi}}$} is an ERC20 or ERC721 token contract. %

Algorithm~\ref{alg_recognizedefiaction} presents the process of \textbf{S-2}. 
\lifter{} takes in a list of 
\mevformula{C}, \mevformula{action$_{\mevsmallformula{type}}$}, and \mevformula{assetTransfers},
and then recognizes the DeFi action (Line 4-11) according to the asset transfer patterns in Table~\ref{table_actsemantics}. Finally, it outputs the recognized DeFi actions in a transaction (Line 14).  
Since different kinds of DeFi actions involve different numbers of asset transfers, we divide them into four categories as follows. %

First, for AddLiquidity and RemoveLiquidity actions that require \mevformula{n} asset transfers, we pick \mevformula{n} asset transfers in a greedy fashion from \mevformula{aTs$_{\mevsmallformula{i}}$} that match the patterns to recognize them (Line 6).  
Second, for Swap and Liquidation actions that require two asset transfers, we pick the first two asset transfers from \mevformula{aTs$_{\mevsmallformula{i}}$} that match the patterns to recognize them (Line 8).
Third, since Rebasing action does not require asset transfers, for contract \mevformula{C$_{\mevsmallformula{DeFi}}$} that conducts the Rebasing action, we check whether \mevformula{C$_{\mevsmallformula{DeFi}}$} implements standard functions defined in ERC20 or ERC721~\cite{frowis2019detecting}.
Fourth, for the other five DeFi actions that require one asset transfer, we pick the first asset transfer from \mevformula{aTs$_{\mevsmallformula{i}}$} that matches the patterns to recognize them (Line 12).

\section{\cluster{}}
\label{sec_actcluster}

\begin{figure}
\centering
  \includegraphics[width=1.0\linewidth]{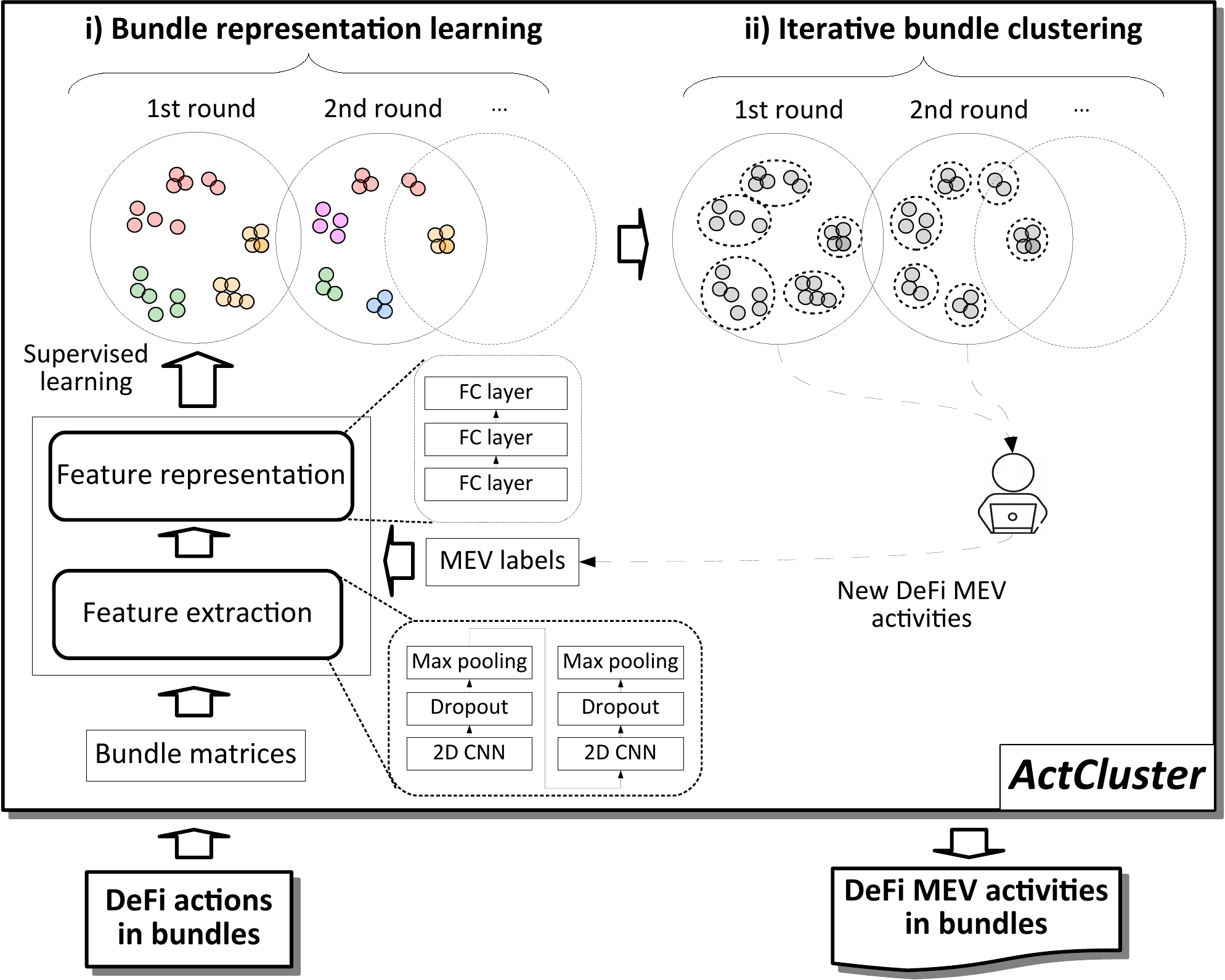}
  \caption{Overview of \cluster{}}
  \vspace{-2pt}
  \label{fig:clustering}
\end{figure}

\cluster{} aims at facilitating analysts to discover DeFi MEV activities in bundles, especially the unknown ones, by analyzing the semantic features involved in the sequences of DeFi actions identified by \lifter{}.
As shown in Fig.~\ref{fig:clustering}, 
it consists of two steps, i.e.,
i) bundle representation learning~(\S \ref{sec_clustersec_2}), which 
maps the raw bundles to their feature vectors in a low-dimensional feature space, and
ii) iterative bundle clustering~(\S \ref{sec_bundleclustering_3}), which
discovers new kinds of DeFi MEV activities via iteratively
clustering feature vectors of bundles.
We repeat the two steps by conducting representation learning with newly discovered DeFi MEV activities in the first step and conducting the iterative clustering analysis in the second step, until we cannot find new DeFi MEV activities.

The design rationale of \cluster{} is fourfold. First, 
manual efforts in inspecting DeFi actions in bundles are required to discover new DeFi MEV activities.
We cluster bundles with similar activities to minimize the manual work.
Second, there is a dilemma in the setting of clustering granularity. Specifically, bigger but sparse clusters may mix bundles containing different DeFi MEV activities together, whereas smaller but denser clusters increase manual efforts in inspecting bundles sampled from each cluster.
We leverage iterative clustering analysis~\cite{liu2019logzip} to address the dilemma, i.e.,
i) it gradually improves the clustering granularity to facilitate the discovery of relatively rare DeFi MEV activities. 
Besides, ii) it reduces the number of clusters that need to be manually inspected through bundle pruning, which iteratively excludes bundles containing known and discovered DeFi MEV activities from the bundle dataset.

Third, conventional clustering algorithms cannot be directly applied to raw bundles
due to bundles' heterogeneous format and hierarchical data structure.
To tackle this problem, we employ representation learning~\cite{bengio2013representation} to automatically extract distinguishable features from raw bundles with the knowledge of all known and discovered DeFi MEV activities.
Unlike feature engineering, which requires rich domain-specific knowledge, representation learning is fully data-driven and task-oriented, obviating considerable manual efforts for data study.
Fourth, in the first round, we conduct the representation learning with three known MEV labels collected from existing studies.
Inspired by previous studies~\cite{xie2021dual,zhu2017iterative} that improve model training's efficiency and performance by scaling up labels and iteratively training with dynamical label updating, 
after each round, we extend new DeFi MEV activities to MEV labels and conduct the representation learning to improve its representation capabilities for DeFi MEV activities in bundles.

\subsection{Bundle representation learning}
\label{sec_clustersec_2}

We map bundles to low-dimensional feature vectors,
based on which the dissimilarity between two bundles can be quantified by the distance between their feature vectors and thus clustering analysis of bundles can be reasonably conducted.

\begin{figure}[t!]
\centering
  \includegraphics[width=0.95\linewidth]{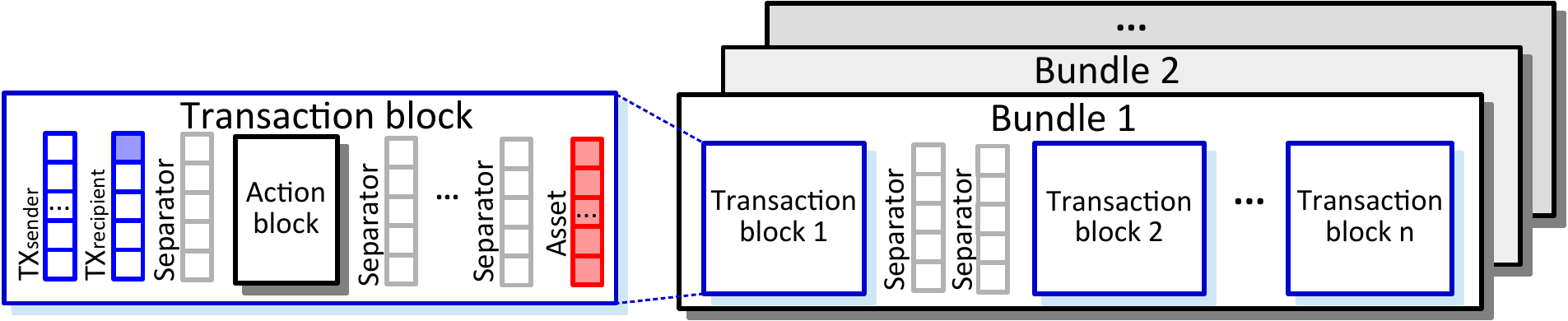}
  \vspace{1pt}
  \caption{Bundle matrix after bundle formatting}
  \label{fig:bundlemetrix}
\end{figure}

\noindent
\textbf{Bundle Formatting.}
Raw bundles are in a heterogeneous format 
since a bundle contains a variable number of transactions,
each of which contains a variable number of DeFi actions.
To facilitate the feature extraction,
we express raw bundles in a unified format,
i.e., a bundle matrix with a fixed shape,
because bundles in the format can be directly processed by convolutional neural network (CNN)~\cite{gu2018recent} in an end-to-end fashion.
Considering that bundles are organized in a hierarchical structure,
we construct a bundle matrix in a bottom-up manner.
Specifically,
we first standardize the description of DeFi actions as ten types of parameterized action blocks,
corresponding to each kind of DeFi action~(\textbf{A1-10} in~\S \ref{sec_notation}).
As shown in Fig.~\ref{fig:bundlemetrix},
each DeFi action in a transaction will be expressed as an action block,
acting as a basic element to describe this transaction.
Sequentially concatenating action blocks corresponding to all DeFi actions within a transaction yields the transaction block that expresses this transaction.
Recall that bundle is essentially a bunch of transactions.
We construct the bundle matrix to express a bundle by
combining transaction blocks corresponding to all transactions within it.
We elaborate more on their constructions 
in Appendix F.

\noindent
\textbf{Feature extraction.}
We extract features from bundle matrices by taking advantage of a CNN. %
The reasons for choosing CNN are threefold: 
i) a bundle can be regarded as a time series
because transactions within it are ordered. The temporal patterns involved in a bundle have been characterized as spatial patterns in our matrix representation of bundles. Thus, our task is suitable for CNN, which is known to be effective and efficient in extracting features from spatial patterns~\cite{gu2018recent}.
ii) typical time series analysis models are not suitable
for our tasks. First, transactions cannot be represented as tokens
as the input of typical models (e.g., Bert and Transformer). Second,
transactions consisting of various actions and parameters are
difficult to be compactly represented as feature vectors with
fixed size as the input of RNN and its variants (e.g., LSTM
and GRU) without information loss.
iii) CNN processes data in parallel and thus is efficient, %
e.g., CNN-based models even achieve state-of-the-art performance in traffic analysis tasks~\cite{sirinam2019triplet}, where samples are represented as time series.
As shown in Fig.~\ref{fig:clustering},
feature extraction is implemented using stacked blocks consisting of a 2D CNN layer, a dropout layer, and a max pooling layer.
Such a network structure facilitates feature extraction because
i) the 2D CNN layer with learnable kernels automatically captures informative features to construct feature maps,
ii) the dropout layer reduces the overfitting risk,
and iii) the max pooling layer downsamples feature maps to highlight the most important feature.
The input is a bundle matrix, and the outputs are feature maps extracted by the last block.

\noindent
\textbf{Feature representation.}
To represent a bundle in a low-dimensional feature space,
we flatten its feature maps obtained via feature extraction and process them with three stacked fully connected (FC) layers.
The output of the last fully connected layer is the low-dimensional feature vector of this bundle.
Models for feature extraction and feature representation are trained by leveraging supervised learning
with the aid of all known MEV labels.
Specifically,
we construct a multi-label classifier based on multilayer perceptron (MLP)~\cite{gardner1998artificial} to classify bundles in the feature space.
We construct the initial MEV labels by
collecting three types of MEV DeFi activities from existing studies~\cite{qin2021quantifying,wang2021cyclic,ferreira2021frontrunner,zhou2021high}, i.e., Sandwich Attack, Cyclic Arbitrage, and Liquidation.
After we discover new DeFi MEV activities in the clustering analysis of each round, we extend our MEV labels with them.
MLP predicts the presence/absence of each label for a bundle. Specifically, 1 (resp. 0) indicates the presence (resp. absence) of a label.
We specify the output layer of MLP as a sigmoid layer so that outputs are normalized in the range of (0, 1). We choose Mean Square Error (MSE) loss~\cite{ucar2021subtab} to quantify the prediction error of MLP. MLP and models for feature extraction and feature representation are jointly trained by minimizing the MSE loss.

\subsection{Iterative bundle clustering}
\label{sec_bundleclustering_3}
Given the feature vectors of bundles,
we characterize the dissimilarity between bundles with the distance between their feature vectors
to facilitate the discovery of new DeFi MEV activities via bundle clustering.
We test five candidate clustering algorithms, including hierarchical clustering~\cite{xu2005survey}, DBSCAN~\cite{xu2005survey}, K-means~\cite{xu2005survey}, Mean Shift~\cite{xu2005survey}, and Birch clustering~\cite{xu2005survey},
and finally cherry-pick DBSCAN for two reasons. First, 
it is more efficient than other algorithms in handling large-scale datasets in our problem. Second, DBSCAN does not need a pre-specified number of clusters.

DBSCAN is a density-based clustering algorithm. Its parameter $\epsilon$ (i.e., the maximum distance between two samples for one to be considered as the other's neighborhood~\cite{xu2005survey}.) adjusts the lower bound of cluster density. A larger $\epsilon$ leads to bigger but sparse clusters, where bundles corresponding to various DeFi MEV activities may be mixed together. 
By contrast, a smaller $\epsilon$ results in smaller but denser clusters, enabling more fine-grained clustering analysis in favor of distinguishing different DeFi MEV activities. However, a side-effect is it substantially increases manual efforts because a smaller $\epsilon$ yields more clusters and we need to manually inspect them to verify whether they contain unseen DeFi MEV activities. To address the dilemma, 
we leverage the iterative clustering analysis by conducting the following steps after representation learning in each round until we cannot discover new DeFi MEV activities:

\noindent
- \textbf{I.} Filter out bundles that only contain DeFi actions that can make up known and discovered DeFi MEV activities.

\noindent
- \textbf{II.}
Group bundles into clusters based on DBSCAN. %

\noindent
- \textbf{III.}
Discover new DeFi MEV activities by sampling one bundle from each cluster and manually inspecting them to determine whether their DeFi MEV activities are new.
To avoid individual bias,
we involve three authors to jointly make a decision,
achieving a consensus on whether a DeFi MEV activity is new.

\noindent
- \textbf{IV.}
Reduce the parameter $\epsilon$ by multiplying it by a decay factor $\eta=0.5$ to improve the resolution of clustering analysis.

Note that the number of bundles for clustering analysis decreases in iterations since bundles associated with discovered DeFi MEV activities are gradually filtered out.
Besides, we reduce the parameter $\epsilon$ of DBSCAN in iterations yielding smaller but
denser clusters.
It enables us to conduct fine-grained clustering analysis for discovering bundles containing unknown DeFi MEV activities.

\section{Evaluation}
\label{sec_evaluation}

We implement \lifter{} and \cluster{} in 7,832 lines of Python code,  maintain an archive Ethereum node, and conduct experiments on a server with an Intel Xeon W-1290 CPU (3.2 GHz, 10 cores), and 128 GB memory to answer four research questions. 
\textbf{RQ1:} How is the performance of \lifter{} in identifying DeFi actions?
\textbf{RQ2:} Does \lifter{} outperform existing techniques with respect to identifying DeFi actions?
\textbf{RQ3:} How many kinds of new DeFi MEV activities does \cluster{} discover? 
\textbf{RQ4:} Does \cluster{} outperform other methods in reducing manual efforts? %

\subsection{Data collection}
\label{sec_datacollection}

\noindent
\textbf{Trace collection.}
We invoke the \texttt{\small debug.traceTransaction()}~\cite{debugTrace2021debugtrace} API of our archive Ethereum node (which is synchronized to the latest state) to
get the transaction execution traces for \lifter{}.

\noindent
\textbf{Bundle collection.}
Since \cluster{} needs the DeFi actions identified by \lifter{} in each bundle, we collect bundles and their transactions by querying the web API~\cite{flashbots2021api} provided by the Flashbots~\cite{flashbot2021bundles}, which displays all bundles and transactions in each bundle mined in Ethereum and relayed by Flashbots. By downloading all bundles from 
the starting date of bundle mechanism
(i.e., Feb. 11, 2021) to Dec. 1, 2022, we collect 6,641,481 bundles and 26,740,394 transactions in total
and form a dataset 
denoted by 
\mevformula{D$_{\mevsmallformula{Bundle}}$}.

\subsection{RQ1: Performance of \lifter{}}
\label{sec_rq1evaluation}

In \textbf{S-1} (\S \ref{sec_marketidentification}), after locating emitted events in $\mathbb{M}$,
\lifter{} recognizes asset transfers involved in DeFi actions, if the asset transfers are logged by the events.
Specifically,
if an event’s parameters contain an asset transfer’s parameters, \lifter{} confirms that the asset transfer is logged by the event.
There are four parameters (i.e., \mevformula{Asset}, \mevformula{From}, \mevformula{To}, and \mevformula{Value}) in each asset transfer.
Hence, \lifter{} can choose a different number of parameters (e.g., the \mevformula{Value} parameter or all four parameters) of an asset transfer and determine whether the parameters are in the event's parameters. 
We evaluate \lifter{} in terms of identifying DeFi actions with the following three configurations and manually determine the number of true positives (TP: a DeFi action is successfully identified), false positives (FP: a non-DeFi action is reported by mistake), and false negatives (FN: a DeFi action is missed) due to the lack of  dataset with ground-truth.
\begin{itemize}[leftmargin=*]
    \item ${c1}$. \lifter{} chooses \mevformula{Value} of each asset transfer and confirms whether the \mevformula{Value} is in the event's parameters. %
    \item ${c2}$. \lifter{} chooses \mevformula{Value} and \mevformula{Asset} of each asset transfer, and confirms whether they are both in the event's parameters. %
    \item ${c3}$. \lifter{} chooses all four parameters of each asset transfer, and confirms whether they are all in the event's parameters.    %
\end{itemize}

We also compute the precision, recall, and f-score~\cite{olson2008advanced}.
Since the number of transactions in \mevformula{D$_{\mevsmallformula{Bundle}}$} is too large ($>$ 10 million), we sample 1,358,122 transactions from \mevformula{D$_{\mevsmallformula{Bundle}}$} for manual inspection and form a dataset denoted by \mevformula{D$_{\mevsmallformula{Trans}}$}. 
To reduce unnecessary manual efforts and mitigate the potential negative effect of human subjectivity on determining TP/FP/FN, we
first de-duplicate transactions having the same execution
traces, because \lifter{} will output identical results for them.
Since the number of transactions after the trace-based de-duplication is still large ($>$ 400,000), we further de-duplicate transactions having the same emitted event sequences (we will evaluate whether such de-duplication will cause errors in the following.). After the event-based de-duplication, 41,090 transactions are left for manual
checking.
Then, six authors manually check these 41,090 transactions. Once we get the TP/FP/FN results for a transaction, all de-duplicated transactions corresponding to this transaction have the same TP/FP/FN results.

Since manual inspection is labor-intensive and might cause errors,
we conduct experiments to evaluate the quality of our TP/FP/FN results.
First, we assess the performance of deduplication %
and provide the confidence level of our results. 
We randomly sample 1000 de-duplicated transactions from the 41,090 transactions, and find that all 1,000 transactions can be de-duplicated.
Note that in relation to the total population ($>$ 1 million), our sample
size has a confidence interval of less than 0.27\%, with 99.9\% confidence.
Second, we compute two statistical measures (i.e., Fleiss’ Kappa~\cite{fleiss} and Krippendorff’s Alpha~\cite{krippendorff}) to assess whether our TP/FP/FN results from different authors reach a consensus.
We randomly sample 500 transactions from the 41,090 transactions, and ask all six authors to report their own results. 
Then, we compute the Fleiss’ Kappa and
Krippendorff’s Alpha to assess the reliability of their manual results.
The results are 0.9884 and
0.9948, respectively, showing that six authors come to an almost perfect agreement.

\begin{table}[]
\centering
\caption{Performance metrics of \lifter{} in identifying DeFi actions with different configurations}
\vspace{1pt}
\resizebox{\linewidth}{!}{
\begin{tabular}{|l|cccccccc|}
\hline
DeFi action type                 & Techniques      & \# Identified & \# TP     & \# FP & \# FN   & Precision & Recall  & F-score \\ \hline
\rowcolor{black!20}             &  \lifter{}$_{c1}$ & 2,156,198     & 2,156,198 & 0     & 27,897  & 100\%     & 98.72\% & 99.36\% \\
\rowcolor{black!20}               & \lifter{}$_{c2}$ & 102,285       & 102,285   & 0     & 2,081,810 & 100\%     & 4.68\%  & 8.95\%  \\
\rowcolor{black!20} \multirow{-3}{*}{Swap}      &  \lifter{}$_{c3}$ & 58,978        & 58,978    & 0     & 2,125,117 & 100\%     & 2.70\%  & 5.26\%  \\ \hline
\rowcolor{black!5}    & \lifter{}$_{c1}$ & 8,056         & 8,056     & 0     & 0       & 100\%     & 100\%   & 100\%   \\
\rowcolor{black!5}               & \lifter{}$_{c2}$ & 45            & 45        & 0     & 8011    & 100\%     & 0.56\%  & 1.11\%  \\
\rowcolor{black!5} \multirow{-3}{*}{AddLiquidity}         & \lifter{}$_{c3}$ & 45            & 45        & 0     & 8011    & 100\%     & 0.56\%  & 1.11\%  \\ \hline
\rowcolor{black!20} & \lifter{}$_{c1}$ & 6,839         & 6,839     & 0     & 0       & 100\%     & 100\%   & 100\%   \\
\rowcolor{black!20}             & \lifter{}$_{c2}$ & 1,198         & 1,198     & 0     & 5,641   & 100\%     & 17,52\% & 29.81\% \\
\rowcolor{black!20} \multirow{-3}{*}{RemoveLiquidity}       & \lifter{}$_{c3}$ & 1,198         & 1,198     & 0     & 5,641   & 100\%     & 17.52\% & 29.81\% \\ \hline
\rowcolor{black!5}     & \lifter{}$_{c1}$ & 1,635         & 1,635     & 0 & 0   & 100\%     & 100\%   & 100\%   \\
\rowcolor{black!5}                                 & \lifter{}$_{c2}$ & 496           & 496       & 0     & 1,139   & 100\%     & 30.34\% & 46.56\% \\
\rowcolor{black!5} \multirow{-3}{*}{Liquidation}     & \lifter{}$_{c3}$ & 496           & 496       & 0     & 1,139   & 100\%     & 30.34\% & 46.55\% \\ \hline
\rowcolor{black!20}     & \lifter{}$_{c1}$ & 16,795        & 16,795    & 0     & 0       & 100\%     & 100\%   & 100\%   \\
\rowcolor{black!20}                                 & \lifter{}$_{c2}$ & 16,795        & 16,795    & 0     & 0       & 100\%     & 100\%   & 100\%   \\
\rowcolor{black!20} \multirow{-3}{*}{NFT-Minting}     & \lifter{}$_{c3}$ & 16,795        & 16,795    & 0     & 0       & 100\%     & 100\%   & 100\%   \\ \hline
\rowcolor{black!5}     & \lifter{}$_{c1}$ & 1,308         & 1,308     & 0     & 0       & 100\%     & 100\%   & 100\%   \\
\rowcolor{black!5}                                 & \lifter{}$_{c2}$ & 1,308         & 1,308     & 0     & 0       & 100\%     & 100\%   & 100\%   \\
\rowcolor{black!5} \multirow{-3}{*}{NFT-Burning}             & \lifter{}$_{c3}$ & 1,308         & 1,308     & 0     & 0       & 100\%     & 100\%   & 100\%   \\ \hline
\rowcolor{black!20}        & \lifter{}$_{c1}$ & 34            & 34        & 0     & 0       & 100\%     & 100\%   & 100\%   \\
 \rowcolor{black!20}                                & \lifter{}$_{c2}$ & 33            & 33        & 0     & 1       & 100\%     & 97.06\% & 98.51\% \\
\rowcolor{black!20} \multirow{-3}{*}{Leverage}             & \lifter{}$_{c3}$ & 33            & 33        & 0     & 1       & 100\%     & 97.06\% & 98.51\% \\ \hline
\rowcolor{black!5} & \lifter{}$_{c1}$ & 684           & 684       & 0     & 0       & 100\%     & 100\%   & 100\%   \\
 \rowcolor{black!5}                                & \lifter{}$_{c2}$ & 191           & 191       & 0     & 493     & 100\%     & 27.92\% & 43.66\% \\
\rowcolor{black!5}\multirow{-3}{*}{Borrowing}      & \lifter{}$_{c3}$ & 191           & 191       & 0     & 493     & 100\%     & 27.92\% & 43.66\% \\ \hline
\rowcolor{black!20}        & \lifter{}$_{c1}$ & 246           & 246       & 0     & 0       & 100\%     & 100\%   & 100\%   \\
\rowcolor{black!20}                                 & \lifter{}$_{c2}$ & 40            & 40        & 0     & 206     & 100\%     & 16.26\% & 27.97\% \\
\rowcolor{black!20} \multirow{-3}{*}{Airdrop}           & \lifter{}$_{c3}$ & 40            & 40        & 0     & 206     & 100\%     & 16.26\% & 27.97\% \\ \hline
\rowcolor{black!5}        & \lifter{}$_{c1}$ & 15            & 15        & 0     & 0       & 100\%     & 100\%   & 100\%   \\
\rowcolor{black!5}                                 & \lifter{}$_{c2}$ & 15            & 15        & 0     & 0       & 100\%     & 100\%   & 100\%   \\
\rowcolor{black!5} \multirow{-3}{*}{Rebasing}         & \lifter{}$_{c3}$ & 15            & 15        & 0     & 0       & 100\%     & 100\%   & 100\%   \\ \hline \hline
\multirow{3}{*}{\textbf{Total}}           & \lifter{}$_{c1}$ & 2,191,810     & 2,191,810 & 0     & 27,897  & 100\%     & 98.74\% & 99.37\% \\
                                 & \lifter{}$_{c2}$ & 122,406       & 122,406   & 0     & 2,097,301 & 100\%     & 5.51\%  & 10.45\% \\
                                 & \lifter{}$_{c3}$ & 79,099        & 79,099    & 0     & 2,140,608 & 100\%     & 3.56\%  & 6.88\%  \\ \hline
\end{tabular}
}
\vspace{1pt}
\label{table_lifterperformance}
\end{table}

\begin{figure}[b]
	\centering
	\includegraphics[width=0.3\textwidth]{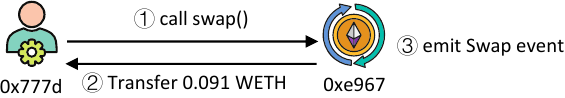}
	\vspace{1pt}
	\caption{An example of \lifter{}$_{c1}$'s false negatives}
	\vspace{-2pt}
	\label{fig_liftererror}
\end{figure}

Table \ref{table_lifterperformance} shows 
the performance of \lifter{} in identifying ten DeFi actions with different configurations.
The third column lists the number of identified DeFi actions for each DeFi action.
The fourth - ninth columns list the number of TPs, FPs, and FNs, and the values of precision, recall, and f-score for each DeFi action.
It shows that
\lifter{}$_{c1}$ (i.e., \lifter{} with the $c1$ configuration)
can effectively identify DeFi actions with nearly 100\% precision and recall. 
However, \lifter{}$_{c1}$ misses 27,897 Swap actions. Manual investigation reveals that traders can receive assets from AMMs at zero cost,
hence there is only one asset transfer in the transaction.
Since \lifter{} identifies a Swap action by matching two asset transfers (\S\ref{sec_actions_formula}),
\lifter{}$_{c1}$ cannot recognize the two asset transfers in transactions, and misses identifying the 27,897 Swap actions in \textbf{S-2} (\S \ref{sec_actions_formula}).
Such cases count for only 1.26\% (27,897/(27,897+2,191,810)) of all DeFi actions, and \lifter{}$_{c1}$ can achieve a 98.74\% recall rate.
Fig.~\ref{fig_liftererror} shows an example, where 
trader \texttt{\small 0x777d} invokes \texttt{\small swap()} of AMM \texttt{\small 0xe967} (i.e., an AMM which supports asset exchanges between \texttt{\small MCC} and \texttt{\small WETH} tokens) in \circlednum{1} in a transaction.
Then AMM \texttt{\small 0xe967} is aware of a difference of \texttt{\small MCC} between AMM \texttt{\small 0xe967}'s token balance and AMM \texttt{\small 0xe967}'s reserve variables of \texttt{\small MCC}.
Please note that AMM \texttt{\small 0xe967}’s reserve variables of \texttt{\small MCC} are stored in AMM \texttt{\small 0xe967}'s contract with aiming of recording AMM \texttt{\small 0xe967}’s token balance of \texttt{\small MCC}.
AMM \texttt{\small 0xe967} considers that the difference of \texttt{\small MCC} is transferred by trader \texttt{\small 0x777d}, and trader \texttt{\small 0x777d} aims to buy \texttt{\small WETH}. 
Hence, AMM \texttt{\small 0xe967} transfers \texttt{\small WETH} to trader \texttt{\small 0x777d} in \circlednum{2} and emits 
a \texttt{\small Swap} event in \circlednum{3}. 
Since there is only one asset transfer in the transaction, \lifter{}$_{c1}$ considers there is no Swap action. %

Unfortunately, \lifter{}$_{c2}$ and \lifter{}$_{c3}$ can only achieve 5.51\% and 3.56\% recall rates.
Manual investigation reveals two reasons for FNs.
First, for the 27,897 Swap actions missed by \lifter{}$_{c1}$,
both \lifter{}$_{c2}$ and \lifter{}$_{c3}$ also missed the 27,897 Swap actions by the same reason.
Second, \lifter{}$_{c2}$ and \lifter{}$_{c3}$ 
missed recognizing 2,069,419 and 2,112,711 DeFi actions, because, due to the configurations of $c2$ and $c3$, the asset transfers for identifying DeFi actions are filtered out. 
For the example in Fig.~\ref{fig_motivating2trace}, \lifter{}$_{c2}$ and \lifter{}$_{c3}$ filter out the two asset transfers in \circlednum{2} and \circlednum{3}, because only the \mevformula{Value} parameters of the two asset transfers (i.e., \mevformula{14,082.22} and \mevformula{500,187}) are logged by the \texttt{\small Swap(}\mevformula{500,187, 14,082.22}\texttt{\small)} event in \circlednum{4}. %
Hence, \lifter{}$_{c2}$ and \lifter{}$_{c3}$ can not identify the corresponding DeFi action, which is matched by the two asset transfers in \circlednum{2} and \circlednum{3}. By contrast, \lifter{}$_{c1}$ can recognize asset transfers involved in DeFi actions (e.g., the two asset transfers in \circlednum{2} and \circlednum{3}), and hence it can identify the corresponding DeFi actions.
Since \lifter{} with the $c1$ configuration achieves nearly 100\% accuracy and significantly outperforms \lifter{} under the other two configurations, i.e, \lifter{}$_{c2}$ and \lifter{}$_{c3}$, we run \lifter{} with the $c1$ configuration for other experiments.

\noindent
\textbf{\textit{Insight.}}
When DeFi developers emit events to announce DeFi actions, we find that
most of them only publish the amount of transferred assets involved in DeFi actions without other parameters of asset transfers (e.g., the type of asset). It may lead to potential risks for traders in interacting with DeFis, because the same events may be triggered by DeFi actions involving different kinds of assets and thus traders will get confused or abused by adversaries.

\noindent\textbf{Answer to RQ1}:
\textit{
\lifter{} can achieve nearly 100\% precision and recall
in identifying ten kinds of DeFi actions.}

\subsection{RQ2: Is \lifter{} superior to others?}
\label{sec_rq2}

We compare \lifter{} with three baseline methods, including two state-of-the-art techniques (i.e., Etherscan~\cite{etherscan2015etherscan} and DeFiRanger~\cite{wu2021defiranger}), and \textsc{EventLifter}, a tool we developed for recognizing DeFi actions from events' arguments, because, as mentioned in \S \ref{sec_preparation}, we find 41 events whose arguments can be leveraged to recognize DeFi actions.  %
We compare their performance in terms of identifying DeFi actions for transactions in \mevformula{D$_{\mevsmallformula{Trans}}$}.
Since Etherscan does not release DeFi action results in its APIs,
we queried their transaction pages to obtain DeFi action results.
Since DeFiRanger is also not available~\cite{wu2021defiranger}, we re-implemented its DeFi action identification approach. 
Note that DeFiRanger~\cite{wu2021defiranger} only identifies AddLiquidity and RemoveLiquidity actions that supply and withdraw single asset with AMMs. For example, if an AddLiquidity action supplies two assets \mevformula{Asset$_1$} and \mevformula{Asset$_2$} to an AMM, DeFiRanger will identify two AddLiquidity actions that supply \mevformula{Asset$_1$} and \mevformula{Asset$_2$} to the AMM, respectively.
We still consider that DeFiRanger identifies the true results, if their results can be combined into the true DeFi actions.

\begin{table}[]
\centering
\caption{Performance metrics of Etherscan, DeFiRanger, and \textsc{EventLifter} in identifying DeFi actions}
\vspace{2pt}
\resizebox{\linewidth}{!}{
\begin{tabular}{|l|cccccccc|}
\hline
DeFi action type                 & Techniques      & \# Identified & \# TP     & \# FP & \# FN   & Precision & Recall  & F-score \\ \hline
\rowcolor{black!20}            & Etherscan  & 1,983,869       & 1,983,869 & 0       & 200,226 & 100\%     & 90.83\% & 95.20\% \\
\rowcolor{black!20}                 & DeFiRanger & 1,760,236       & 1,356,586 & 403,650 & 827,509 & 77.07\%   & 62.11\% & 68.79\% \\
\rowcolor{black!20} \multirow{-3}{*}{Swap}        & \textsc{EventLifter} &  102,285       &  102,285 & 0 & 2,081,810 & 100\%   & 4.68\% &  9.16\% \\ \hline
\rowcolor{black!5}    & Etherscan  & 4,964         & 4,964   & 0       & 3,092   & 100\%     & 61.62\% & 76.25\% \\
\rowcolor{black!5}                & DeFiRanger & 24,234        & 4,116   & 20,118   & 7,237   & 16.98\%    & 36.25\% & 23.13\% \\ 
\rowcolor{black!5} \multirow{-3}{*}{AddLiquidity}    & \textsc{EventLifter} & 45       & 45 & 0 & 8,011 & 100\%   & 0.56\% & 1.11\% \\                \hline
\rowcolor{black!20} & Etherscan  & 2,629         & 2,629   & 0       & 4,210   & 100\%     & 38.44\% & 55.53\% \\
\rowcolor{black!20}                & DeFiRanger & 12,289         & 1,143     & 11,146   & 8,270   & 9.3\%    & 12.14\%  & 10.53\%  \\
\rowcolor{black!20} \multirow{-3}{*}{RemoveLiquidity}    & \textsc{EventLifter} & 1,198       & 1,198 & 0 & 5,641 & 100\%   & 17.52\% &  29.81\% \\
                \hline
\rowcolor{black!5}     & Etherscan  & 527           & 527     & 0       & 1,108     & 100\%     & 32.23\% & 48.75\% \\
\rowcolor{black!5}                & DeFiRanger & -             & -       & -       & -       & -         & -       & -       \\
\rowcolor{black!5} \multirow{-3}{*}{Liquidation}   & \textsc{EventLifter} & 496       & 496 & 0 & 1,139 & 100\%   &  30.34\% & 46.55\% \\                \hline
                
\rowcolor{black!20} & Etherscan  & 12,532         & 12,532   & 0       & 4,263   & 100\%     & 74.62\% & 85.46\% \\
\rowcolor{black!20}                & DeFiRanger & -           & -     & -       & -     & -     & - & - \\ 
\rowcolor{black!20} \multirow{-3}{*}{NFT-Minting}    & \textsc{EventLifter} & -       & - & - & - & -   & - & - \\                \hline
\rowcolor{black!5}     & Etherscan  & 1,190           & 1,190     & 0       & 118     & 100\%     & 90.98\% & 95.28\% \\
 \rowcolor{black!5}                & DeFiRanger & -             & -       & -       & -       & -         & -       & -       \\
\rowcolor{black!5}  \multirow{-3}{*}{NFT-Burning}                & \textsc{EventLifter} & -       & - & - & - & -   & - & - \\                \hline

\rowcolor{black!20} & Etherscan  & -           & -     & -       & -     & -     & - & - \\
\rowcolor{black!20}               & DeFiRanger & -           & -     & -       & -     & -     & - & -  \\
\rowcolor{black!20}\multirow{-3}{*}{Leverage}        & \textsc{EventLifter} & 33       & 33 & 0 & 1 & 100\%   & 97.06\% &  98.51\% \\                \hline
\rowcolor{black!5}     & Etherscan  & 141           & 141     & 0       & 543     & 100\%     & 20.61\% & 34.18\% \\
\rowcolor{black!5}                & DeFiRanger & -             & -       & -       & -       & -         & -       & -       \\
\rowcolor{black!5}\multirow{-3}{*}{Borrowing}      & \textsc{EventLifter} & 191       & 191 & 0 & 493 & 100\%   & 27.92\% & 43.66\% \\                \hline
                
\rowcolor{black!20} & Etherscan  & -           & -     & -       & -     & -     & - & - \\
\rowcolor{black!20}                & DeFiRanger & -           & -     & -       & -     & -     & - & - \\ 
\rowcolor{black!20} \multirow{-3}{*}{Airdrop}                & \textsc{EventLifter} & 40       & 40 & 0 & 206 & 100\%   &  16.26\% & 27.97\% \\                \hline

\rowcolor{black!5}    & Etherscan  & -           & -     & -       & -     & -     & - & - \\
\rowcolor{black!5}                & DeFiRanger & -             & -       & -       & -       & -         & -       & -       \\
\rowcolor{black!5} \multirow{-3}{*}{Rebasing}     & \textsc{EventLifter} & 15       & 15 & 0 & 0 & 100\%   & 100\% & 100\% \\ \midrule \hline
\multirow{3}{*}{\textbf{Total}}           & Etherscan  & 2,005,852       & 2,005,852 & 0       & 213,560 & 100\%     & 90.38\% & 94.95\% \\
                & DeFiRanger & 1,796,759       & 1,361,845 & 434,914 & 843,016 & 75.79\%   & 61.77\% & 68.06\% \\
                & \textsc{EventLifter} &  104,303       & 104,303 & 0 & 2,097,301 & 100\%   &  4.74\% & 9.05\% \\                \hline
\end{tabular}
}
\label{table_comparison}
\end{table}

Table \ref{table_comparison} shows %
the results of Etherscan, DeFiRanger, and \textsc{EventLifter}.
The third 
- ninth columns list the number of identified DeFi actions, TPs, FPs, and FNs, and the values of precision, recall, and f-score for different kinds of DeFi actions.
We next present reasons why three baseline techniques generate FP and FN cases.

\noindent\textbf{Etherscan.} Etherscan achieved 100\% precision for the identification of 7 kinds of DeFi actions but generated incomplete results.
For example, Etherscan missed identifying a Swap action 
in the transaction~\cite{experiments2021etherscanmmiss}, where
the trader exchanges \texttt{\small COMP} for \texttt{\small WETH} with the AMM \texttt{\small 0xba12}. Both \lifter{} and DeFiRanger can correctly identify this Swap action.
Since Etherscan does not disclose how they identify DeFi actions, we cannot know why Etherscan failed.%

\noindent\textbf{DeFiRanger.} 
DeFiRanger led to both incomplete and inaccurate results in identifying three kinds of actions. 
The root causes are twofold:
i) DeFiRanger identifies DeFi actions by matching ERC20 token transfers~\cite{wu2021defiranger}, and thus it cannot identify DeFi actions involving Ether transfers. 
ii) Heuristics, defined by DeFiRanger~\cite{wu2021defiranger}, are inaccurate. For example, 
DeFiRanger identifies Swap actions by only matching the first two token transfers, and in the two matched token transfers one account receives and sends out different assets.
However, token transfers, which are irrelevant to DeFi actions, can also satisfy these heuristics.
Hence, DeFiRanger will wrongly match the irrelevant token transfers and report wrong DeFi actions.
For the example in Fig.~\ref{fig_motivating2trace}, DeFiRanger identifies an incorrect Swap action by matching two irrelevant token transfers in \circlednum{1} and \circlednum{5}.

\noindent
\textbf{\textsc{EventLifter}.}
\textsc{EventLifter} only achieved 4.74\% recall rate and missed identifying 2,097,316 DeFi actions.
It shows that the 41 kinds of events 
only count for a small proportion (i.e., 4.74\%) of all emitted events whose signatures are in $\mathbb{M}$. %

\noindent\textbf{Answer to RQ2}:
\textit{
\lifter{} can significantly outperform the state-of-the-art techniques, other baseline methods, and two variants of \lifter{}
in identifying DeFi actions.}

\subsection{RQ3: DeFi MEV activities discovery}
\label{sec_rq3evaluation}

The representation learning of \cluster{} leverages three initial MEV labels, i.e., Sandwich Attack, Cyclic Arbitrage, and Liquidation, 
to map each bundle matrix to the low-dimensional feature space in the first round~(\S \ref{sec_clustersec_2}). 
We generate the initial MEV labels for each bundle in \mevformula{D$_{\mevsmallformula{Bundle}}$}
by using heuristics from Qin et al.~\cite{qin2021quantifying}.
As a result, 813,188, 1,334,207, and 14,263 bundles are labeled as Sandwich Attack, Cyclic Arbitrage, and Liquidation, respectively.
Besides, two parameters are used in iterative bundle clustering (\S\ref{sec_bundleclustering_3}) of \cluster{}, i.e., $\epsilon$, which is used to adjust the lower bound of cluster density in DBSCAN~\cite{xu2005survey}, and $\eta$, which is the delay factor of $\epsilon$.
The values of $\epsilon$ and $\eta$ are selected by grid search~\cite{wang2015deep} with the target of minimizing required manual efforts (i.e., the amount of bundles manually inspected) to discover MEV activities.
Specifically, we first make a set of candidate values for $\epsilon$ and $\eta$, and then perform the iterative bundle clustering (\S\ref{sec_bundleclustering_3}) with each pair of parameters on a small set of (i.e., 5,000) bundles in \mevformula{D$_{\mevsmallformula{Bundle}}$}. 
Finally, we compare the amount of bundles manually inspected in iterative bundle clustering (\S\ref{sec_bundleclustering_3}), and determine 16 and 0.5 for $\epsilon$ and $\eta$, respectively.

We train on all our data (i.e., 6,641,481 bundles in \mevformula{D$_{\mevsmallformula{Bundle}}$}) with MEV labels.
After each round in \cluster{}, we add the newly discovered MEV activities into the MEV labels, and conduct the representation learning (\S\ref{sec_clustersec_2}) of the next round with the extended MEV labels. 
After repeating the steps of \cluster{} (\S\ref{sec_actcluster}) four rounds and analyzing 2,035 bundles manually, we discover 17 new kinds of DeFi MEV activities summarized in Table~\ref{table_11mevactsdest}.
We describe one as follows, and introduce the rest in Appendix E.

\begin{table*}[]
\centering
\caption{Descriptions for the 17 kinds of new DeFi MEV activities in bundles}
\vspace{2pt}
\resizebox{\linewidth}{!}{
\begin{tabular}{|l|l|}
\hline
DeFi MEV activity &
  Description \\ \hline
\rowcolor{black!20} Swap Backrun Arbitrage &
  \begin{tabular}[c]{@{}l@{}}
  On the same AMM, the arbitrageur just backruns another trader's Swap action by a Swap action, and earns profits from the pulled-up price.
  \end{tabular} \\ %
 Liquidity Backrun Arbitrage &
  \begin{tabular}[c]{@{}l@{}} On the same AMM, the arbitrageur backruns another trader's AddLiquidity/RemoveLiquidity action by a Swap action, and earns profits from the pulled-up price.\end{tabular} \\ %
\rowcolor{black!20} Liquidity Sandwich Arbitrage &
  \begin{tabular}[c]{@{}l@{}}On the same AMM, the arbitrageur frontruns and backruns another trader's Swap action by AddLiquidity and RemoveLiquidity actions, and earns profits from\\ the trader's exchange fee.\end{tabular} \\ %
Multi-layered Burger Arbitrage &
  \begin{tabular}[c]{@{}l@{}}On the same AMM, the arbitrageur frontruns and backruns other traders' Swap actions by Swap actions, and earns profits from the pulled-up price.\end{tabular} \\ %
\rowcolor{black!20} Liquidity-swap Trade &
  \begin{tabular}[c]{@{}l@{}}On the same AMM, the trader both performs a Swap action, and performs the AddLiquidity or RemoveLiquidity actions. The trader aims to supply, withdraw,\\ or trade assets at the expected prices.\end{tabular} \\ %
Partial Cyclic Arbitrage &
  \begin{tabular}[c]{@{}l@{}}The arbitrageur performs Swap actions among AMMs to earn profits from the unbalanced prices, and part of the Swap actions can form a loop one by one.\end{tabular} \\ %
\rowcolor{black!20} Backrun Cyclic Arbitrage &
  \begin{tabular}[c]{@{}l@{}}The arbitrageur backruns another trader's Swap/AddLiquidity/RemoveLiquidity action, and performs Cyclic Arbitrage to earn profits from the unbalanced prices.\end{tabular} \\ %
Hybrid Arbitrage &
  \begin{tabular}[c]{@{}l@{}}There are at least two kinds of MEV activities of known MEV activities in a bundle. There exists a transaction contained in all these MEV activities. 
  \end{tabular} \\ %
\rowcolor{black!20} Failed Arbitrage &
  \begin{tabular}[c]{@{}l@{}}The arbitrageur suffers the financial loss, when the arbitrageur aims to obtain profits by Sandwich Attack or Cyclic Arbitrage activities.\end{tabular} \\ %
Non-cyclic Swap Trade &
  \begin{tabular}[c]{@{}l@{}}The trader only performs the non-cyclic Swap actions, and aims to trade on the AMMs at the expected prices.\end{tabular} \\ %
\rowcolor{black!20} Rebasing Backrun Arbitrage &
  \begin{tabular}[c]{@{}l@{}}The arbitrageur backruns a Rebasing action by a Swap action, and earns profits from the price differences of the Rebase token.\end{tabular} \\ %
Airdrop-swap Trade &
  \begin{tabular}[c]{@{}l@{}}The trader first claims the airdrop rewards, then sells the received rewards to an AMM by a Swap action.\end{tabular} \\ %
\rowcolor{black!20} Bulk NFT-Minting &
  \begin{tabular}[c]{@{}l@{}}The NFT contract mints NFTs in bulk, and it aims to increase the maintained NFTs at the expected blockchain state.\end{tabular} \\ %
NFT Reforging &
  \begin{tabular}[c]{@{}l@{}}The NFT contract reforges an NFT to update the asset represented by the NFT. \end{tabular} \\ %
\rowcolor{black!20} Airdrop Claiming &
  \begin{tabular}[c]{@{}l@{}}The trader only claims and receives airdrop rewards.\end{tabular} \\ %
NFT-Minting-swap Trade &
  \begin{tabular}[c]{@{}l@{}}The trader first receives an NFT minted by NFT contract, then sells the minted NFT to an AMM by a Swap action.\end{tabular} \\
\rowcolor{black!20} Loan-powered Arbitrage &
  \begin{tabular}[c]{@{}l@{}}The arbitrageur loans assets from Lending under the over/under-collateral deposit, then uses the loaned assets to conduct MEV activities, e.g., Cyclic Arbitrage.\end{tabular} \\ \hline
\end{tabular}
}
\label{table_11mevactsdest}
\vspace{4pt}
\end{table*}

\begin{figure}[t!]
	\centering
	\includegraphics[width=0.33\textwidth]{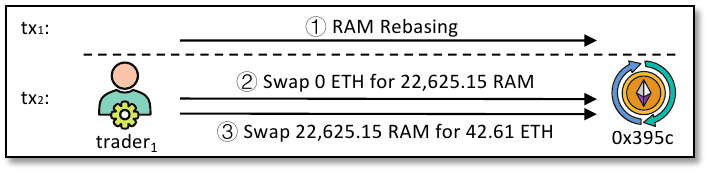}
	\vspace{1pt}
	\caption{An example of Rebasing Backrun Arbitrage}
	\label{fig_rbaexample}
\end{figure}

\noindent
$\bullet$
\textbf{Rebasing Backrun Arbitrage (RBA).}
It involves two transactions in a bundle. 
The former executes a
Rebasing action, which causes a difference
between the AMM's Rebase token balance and the AMM's reserve variables. The Rebase token balance is stored in the contract of Rebase token, and the reserve variables are stored in the contract of AMM with recording the AMM's Rebase token balance.
The latter executes a Swap action to trade the Rebase token and obtains profits from the difference of Rebase token. %
For example, Fig.~\ref{fig_rbaexample} shows the two transactions in the first bundle of the 12,147,015 block. In the first transaction, the \texttt{\small RAM} token executes a Rebasing action, and it causes 
the \texttt{\small RAM} token balance of the AMM \texttt{\small 0x395c} to increase by 22,625.15.
However, the AMM \texttt{\small 0x395c} still uses its old RAM token balance (i.e., the reserve variables) before the Rebasing action to calculate how much the traders should pay~\cite{Adams2020UniswapVC}.
In the second transaction, after giving the trader the 22,625.15 \texttt{\small RAM}, the AMM \texttt{\small 0x395c} finds its \texttt{\small RAM} token balance does not decrease. Hence, the trader$_1$ does not need to pay for \texttt{\small ETH}.
The trader$_1$ then swaps the 22,625.15 \texttt{\small RAM} for 42.61 \texttt{\small ETH} to earn profits of 42.61 \texttt{\small ETH}.

We also evaluate whether our results can generalize to new MEV activities. %
We first train our model on a small set of (i.e., 20,000) bundles in \mevformula{D$_{\mevsmallformula{Bundle}}$},
and then evaluate our trained model on different validation sets~\cite{noroozi2017representation} (i.e., other bundle sets randomly sampled from \mevformula{D$_{\mevsmallformula{Bundle}}$}).
It shows that our trained model achieves similar accuracy (difference < 5\% ) in classifying MEV labels on different validation sets. It means that our model could generalize to different sets of bundles.
As new DeFi MEV activities can cause concept drift~\cite{gama2014survey} in bundles (an open problem in machine learning) and might affect the accuracy of our model, Users can retrain~\cite{gama2014survey} our model with new MEV activities. %
We evaluated the retraining cost of our model, and the result shows that it is reasonable (Appendix J).

\noindent\textbf{Answer to RQ3}: \textit{\cluster{} empowers us to discover 17 new kinds of DeFi MEV activities in bundles. Besides, our results can generalize to new types of DeFi MEV activities.}

\subsection{RQ4: Is \cluster{} superior to others?}
\label{sec_rq4}

To evaluate how much manual effort can be reduced by  \cluster{}.
We compare it with three baseline strategies. 
It is worth noting that the three baseline strategies are selected with ablating components of \cluster{}.
Hence, by comparing with the three baseline strategies, we can also evaluate to which extent the components of \cluster{} benefit the procedure of MEV activity discovery.

\noindent
$\bullet$
\textbf{\cluster{}$^-$.} It ablates the updating of labels of newly discovered MEV activities in model training of bundle representation learning (\S\ref{sec_clustersec_2}).
We only conduct the representation learning (\S \ref{sec_clustersec_2}) with the initial three MEV labels (\S \ref{sec_clustersec_2}), and conduct the iterative clustering analysis by repeating the same four steps in~\S \ref{sec_bundleclustering_3}, until we find all 17 DeFi MEV activities.

\noindent
$\bullet$
\textbf{Intuitive clustering analysis.} It ablates both the iterative bundle clustering (\S\ref{sec_bundleclustering_3}), and the updating of labels of newly discovered MEV activities in model training of bundle representation learning (\S\ref{sec_clustersec_2}).
We apply the DBSCAN algorithm to all bundles in \mevformula{D$_{\mevsmallformula{Bundle}}$} once to find different kinds of DeFi MEV activities. Then, we sample one bundle from each cluster, and determine whether it contains new DeFi MEV activities.
Since we aim to compare \cluster{} with the best performance of the intuitive clustering analysis, we adjust the $\epsilon$ parameter of the DBSCAN clustering algorithm to find all 17 kinds of DeFi MEV activities.

\noindent
$\bullet$
\textbf{Random sampling analysis.} It ablates the whole process of \cluster{}.
We sample one bundle from \mevformula{D$_{\mevsmallformula{Bundle}}$} randomly, and determine whether it contains discovered DeFi MEV activities.
If that is the case, we exclude all bundles containing the corresponding DeFi MEV activities from \mevformula{D$_{\mevsmallformula{Bundle}}$}. Note that the excluded bundles only contain DeFi actions that can
form the corresponding DeFi MEV activities.
We repeat the random sampling analysis until we find all 17 kinds of new DeFi MEV activities.

For each strategy, we record the number of bundles to be inspected for discovering all 17 kinds of new DeFi MEV activities.
Our experimental results show that \cluster{}$^-$, intuitive clustering analysis, and random sampling analysis, require us to manually analyze 2,874, 108,962, and 176,255 bundles, respectively.
Compared to them, \cluster{} can reduce 29.2\%, 98.1\% and 98.8\% of manual efforts for discovering DeFi MEV activities, respectively.

\noindent\textbf{Answer to RQ4}: \textit{\cluster{} outperforms three baseline strategies
in reducing manual efforts during discovering DeFi MEV activities.}

\section{Applications of our approach}
\label{sec_applications}

We use three applications to demonstrate usages of our approach (i.e., \lifter{} and \cluster{}), including enhancing relays' MEV countermeasures (\S\ref{sec_hunter}), evaluating forking and reorg risks caused by MEV activities in bundles (\S\ref{sec_consensus_security}), and evaluating the impact of MEV activities in bundles on blockchain users' economic security (\S\ref{sec_network_security}). Moreover, we discuss three feasible usages of our approach, supported by experimental results and observations (\S\ref{sec_other_applications}).

\subsection{Enhancing MEV countermeasures in relays}
\label{sec_hunter}

As the most popular platforms implementing MEV countermeasures in practice~\cite{yang2022sok}, 
relays that distribute bundles to miners/validators can filter out bundles including known MEV activities~\cite{mevrelay2021list} (e.g., relays~\cite{bloxroute2021bundles,MEVBlocker} block sandwich attacks).
However, 
these relays~\cite{bloxroute2021bundles,MEVBlocker} rely on handcrafted heuristics~\cite{mevinspect2021} to detect and filter out the bundles containing known MEV activities.
Hence, these relays can fail to counter bundles only containing unknown MEV activities because these bundles can fail heuristics of these relays.
We develop a tool named \hunter{} based on our approach to enhance relays to counter bundles containing new MEV activities. 
Specifically,
\hunter{} takes in a bundle of transactions as input, and identifies the kinds of MEV activities (including known and our newly discovered MEV activities) exist in the bundle.
For each transaction in the bundle, \hunter{} utilizes \lifter{} to recognize DeFi actions in it.
Besides, for each kind of MEV activity discovered by \cluster{}, we summarize heuristics to identify it like others~\cite{qin2021quantifying,ferreira2021frontrunner,weintraub2022flash}.
For each kind of MEV activity, the heuristics describe DeFi actions that a bundle arbitrageur has to perform to accomplish the corresponding MEV activity.
By checking whether the DeFi actions in the bundle satisfy our summarized heuristics,
\hunter{} identifies MEV activities in the bundle.

To evaluate how \hunter{} enhance MEV countermeasures in relays, we use it to inspect MEV activities for all bundles in \mevformula{D$_{\mevsmallformula{Bundle}}$}.
The experimental results show that
31.81\% (2,112,344/6,641,481) bundles contain known MEV activities (i.e., Sandwich Attach, Cyclic Arbitrage, and Liquidation), and 53.12\% (3,527,655/6,641,481) bundles contain our newly discovered DeFi MEV activities.
Among the 3,527,655 bundles, 3,182,363 bundles only contain new DeFi MEV activities.
The experimental results indicate that, \hunter{} can enhance relays to additionally identify 3,182,363 (47.92\%) bundles only containing the 17 kinds of new MEV activities.
We further investigate new MEV activities in bundles, e.g., the number of contracts directly invoked by the EOA account to perform new MEV activities. Our empirical results show that new MEV activities are commonly used in bundles (cf. Appendix G for details).

\noindent\textbf{Summary}: \textit{Our approach can enhance MEV countermeasures in relays to discover more MEV activities in bundles, and filter out more bundles (relayed by them) containing MEV activities.}

\subsection{Evaluating forking and reorg risks caused by bundle MEV activities}
\label{sec_consensus_security}

Prior studies~\cite{Daian2020flash,qin2021quantifying,liu2022empirical,zhou2021just} report that financially rational miners are incentivized to deliberately fork and reorganize the blockchain to gather revenues from MEV activities.
Hence,
we evaluate forking and reorg risks in blockchain consensus security caused by known and new MEV activities in bundles (i.e., bundle MEV activities) by measuring how many revenues miners can gather from bundle MEV activities.
Our methodology for determining miners' revenues from known and new MEV activities in bundles in \mevformula{D$_{\mevsmallformula{Bundle}}$} involves two steps. First, we recognize bundles containing known and new MEV activities by using \hunter{} as discussed in \S\ref{sec_hunter}.
Second, following methods in~\cite{liu2022empirical}, we determine miners' revenues from bundle MEV activities by using miners' revenues from corresponding bundles.
Miners' revenues from bundles consist of two parts: i) gas fees for transactions in bundles, and ii) Ether transfers to miners in bundles (both of them are publicly available through the web API~\cite{flashbots2021api}).
To facilitate analysis, we combine miners' revenues from bundle MEV activities per block, and form a dataset denoted by \mevformula{D$_{\mevsmallformula{Revenue}}$}, because miners' revenues from bundle MEV activities contained in the same block will cumulatively incentivize miners to fork and reorganize the blockchain.
As a result, miners receive revenues from bundle MEV activities in 1,791,891 blocks. In block 14,953,916, miners received the highest revenues from bundle MEV activities as 1,584.4 Ether (792.2 times the block reward).

\begin{figure}
	\centering
	\includegraphics[width=0.98\linewidth]{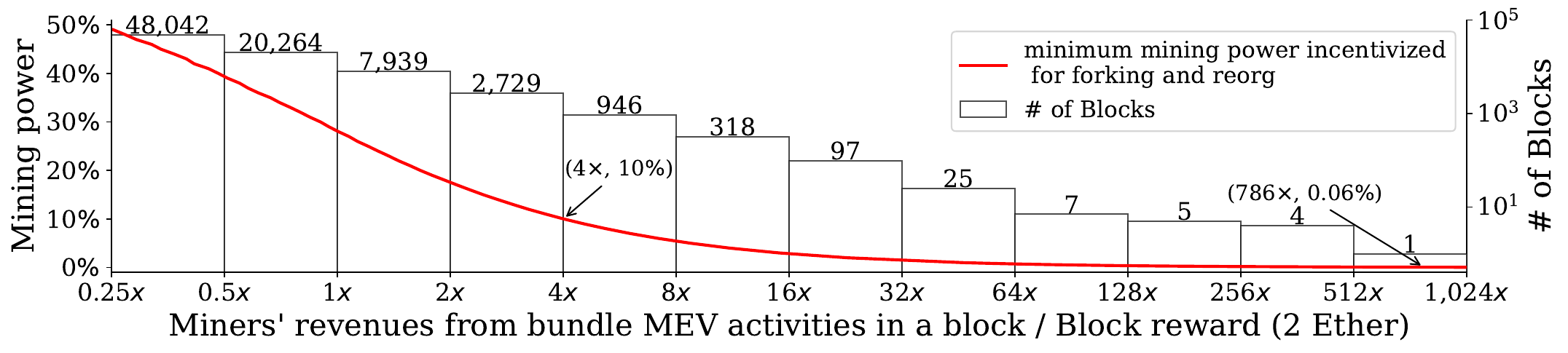}
        \caption{The minimum mining power of a miner incentivized for forking and reorg to gather miners' revenues from a block, and the distribution of miners' revenues in blocks.}	
	\label{fig_consensus_security}
\end{figure}

To further investigate how bundle MEV activities incentivize miners to fork and reorganize the blockchain,
by adapting the MDP framework~\cite{qin2021quantifying},
we quantify the minimum mining power of miners incentivized to fork and reorganize the blockchain for gathering revenues in \mevformula{D$_{\mevsmallformula{Revenue}}$}. 
Specifically, the MDP framework employs a Markov Decision Process~\cite{zhou2021just} for miners to identify whether to fork and reorganize the blockchain or not, with a given mining power on various revenues. 
The results are shown in Fig.~\ref{fig_consensus_security}, where each point (\mevformula{x}, \mevformula{y}) in the red line indicates that, miners' revenues from bundle MEV activities in a block (which are \mevformula{x} times the block reward) will incentive miners with no less than \mevformula{y} mining power to fork and reorganize the blockchain for gathering the revenues.
Besides, in Fig.~\ref{fig_consensus_security}, we display the distribution of miners' revenues from bundle MEV activities in blocks in \mevformula{D$_{\mevsmallformula{Revenue}}$} with binning in twelve intervals.
Fig.~\ref{fig_consensus_security} shows that 1,403 blocks incentivize miners with no less than 10\% mining power to fork and reorganize the blockchain.
Moreover, the miners' revenues from bundle MEV activities in block 14,953,916 can incentivize a miner with only 0.06\% mining power to fork and reorganize the blockchain, highlighting the severe of forking and reorg risks caused by MEV activities in bundles.

Ethereum changed its consensus mechanism from PoW to PoS in
September 2022~\cite{ethereum2022merge}, and the new PoS consensus mechanism is under the forking and reorg risks undertaken by validators~\cite{d2022no}. Besides, several studies~\cite{neu2022two,neuder2021low} propose various attacks to decrease the cost for launching forking and reorg for Ethereum blockchain.
Considering that validators collect the same revenues from bundle MEV activities as miners~\cite{flashbot2021bundles,mevboost2022}, we believe that bundle MEV activities still endanger the consensus security in the context of PoS by incentivizing validators to fork and reorganize the blockchain.

\noindent\textbf{Summary}: \textit{Bundle MEV activities endanger the consensus security by incentivizing miners/validators to fork and reorganize the blockchain for gathering revenues from bundle MEV activities.}

\subsection{Evaluating impact of bundle MEV activities on blockchain users' economic security}
\label{sec_network_security}

To explore the impact of bundle MEV activities on blockchain users' economic security,
we use the Granger causality test~\cite{toda1995statistical,faes2017multiscale} to examine the range of later blocks in which users' transactions are delayed due to bundle MEV activities in prior blocks.
In the context of PoW,
delayed waiting time is one of the major economic security issues for users caused by MEV activities~\cite{qin2021quantifying,liu2022empirical}. For instance, it prolongs users' transactions to be exposed to arbitrageurs, 
thereby enhancing arbitrageurs to design and engage in more profitable MEV activities (e.g., Sandwich Attack and Cyclic Arbitrage).
The Granger causality test is a statistical hypothesis test
to determine whether the changes in one time series cause changes in another time series, and it is widely employed in the fields of economics, political science, and epidemiology~\cite{toda1995statistical,faes2017multiscale}.
We capture the two parts of data examined for our Granger causality test in the following.

\noindent
\textbf{Transaction waiting times}.
We define the waiting time of a transaction as the duration that the transaction remains in mempools of miners/validators before being submitted to blockchain.
To capture transaction waiting times, we utilize the three-month waiting time dataset (from Jul. 20, 2021 to Oct. 27, 2021) released by~\cite{liu2022empirical}.  
Additionally, we obtained a nine-day waiting time dataset for transactions from Mar. 14, 2023 to Mar. 22, 2023 by implementing the same methods as \cite{liu2022empirical} (cf. Appendix K for details). 
We use median values of transaction waiting times in each block to account for the variation of transaction waiting times in blocks, which is more tolerant of outliers than the mean and standard deviation~\cite{liu2022empirical}.
Finally,
we combine waiting times from two time periods to form a new dataset denoted by \mevformula{D$_{\mevsmallformula{Waiting}}$}, which includes the 25th, 50th, and 75th quartiles of waiting times per block (where the 25th, 50th, and 75th quartiles of waiting times are sorted in ascending order).

\noindent
\textbf{Extractable value}.
Following the methods in~\cite{liu2022empirical}, we estimate the extractable value of bundle MEV activities by using revenues of miners/validators from bundle MEV activities (\S\ref{sec_consensus_security}).
It benefits us in estimating the extractable value of bundle MEV activities even if assets in MEV activities do not have price information for calculating the extractable value~\cite{liu2022empirical,qin2021quantifying}.
Please note that \mevformula{D$_{\mevsmallformula{Waiting}}$} includes waiting times for two periods.
For the first period (i.e., from Jul. 20, 2021 to Oct. 27, 2021), 
we obtain the extractable value of bundle MEV activities in blocks by using corresponding results in \mevformula{D$_{\mevsmallformula{Revenue}}$} (\S \ref{sec_consensus_security}). 
Moreover, to obtain the extractable value of bundle MEV activities in blocks for the second period (i.e., from Mar. 14, 2023 to Mar. 22, 2023), we first capture bundles from the web API~\cite{flashbots2021api}, and then use the methods in~\S \ref{sec_consensus_security} to obtain the extractable value of bundle MEV activities in blocks. 
Finally, we combine two parts of results to form a new dataset denoted by \mevformula{D$_{\mevsmallformula{Value}}$}.

\begin{figure}
	\centering
	\includegraphics[width=0.98\linewidth]{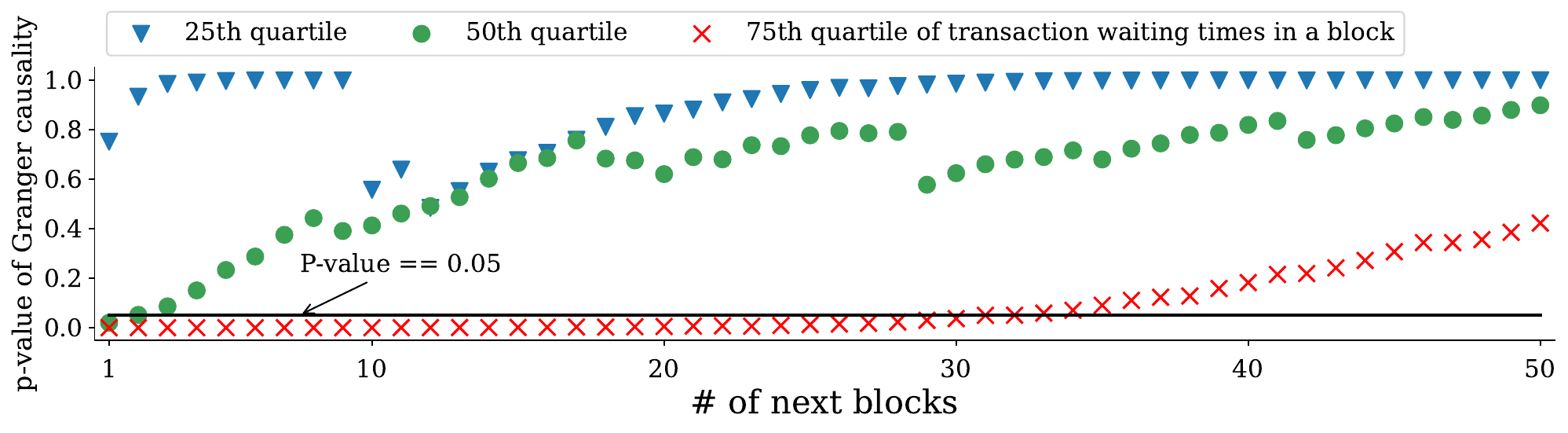}
        \caption{Bundle MEV activities in blocks cause the delayed transaction waiting times in the next $x$-th block, if the corresponding p-value is smaller than 0.05.}
	\label{fig_network_security}
\end{figure}

Our Granger causality test serves to examine the range of later blocks in which the waiting times of users' transactions are prolonged by prior bundle MEV activities. 
Inspired by the lagged Granger causality analysis~\cite{faes2017multiscale},
we achieve this purpose by lagging \mevformula{x} block (1 $\leq$ \mevformula{x} $\leq$ 50) of data in \mevformula{D$_{\mevsmallformula{Value}}$}, and conducting the Granger causality test for the data in \mevformula{D$_{\mevsmallformula{Waiting}}$} and \mevformula{x}-lagged data in \mevformula{D$_{\mevsmallformula{Value}}$}.
If the p-value of corresponding Granger causality test is smaller than 0.05, we confirm that bundle MEV activities cause the increase of corresponding transaction waiting times in the later \mevformula{x}-th block (at the 5\% level of significance~\cite{toda1995statistical,faes2017multiscale}).
Our results are illustrated in Fig.~\ref{fig_network_security}. 
It shows that bundle MEV activities cause the increase of transaction waiting times at the 25th, 50th, and 75th quartiles in next 0, 2, and 30 blocks, respectively.  
Hence, it indicates that bundle MEV activities in blocks cause delayed waiting times of transactions in later blocks.
Please note that, since miners/validators prioritize transactions with higher fees~\cite{pacheco2022my, liu2022empirical}, transactions with lower fees are commonly positioned in the back of the block~\cite{pacheco2022my, liu2022empirical}.
Hence, our results also indicate that the further back in the block a transaction will be positioned (i.e., the transaction has a lower fee), the more continuous delay that bundle MEV activities cause on its waiting time.
We validate our results by measuring the correlation between bundle MEV activities and transaction waiting times via correlation tests (e.g., Spearman~\cite{myers2004spearman}).
Our results show that Spearman coefficients between bundle MEV activities in blocks in \mevformula{D$_{\mevsmallformula{Value}}$} and transaction waiting times at the 25th, 50th, and 75th quartiles in blocks in \mevformula{D$_{\mevsmallformula{Waiting}}$} are 0.230, 0.233, and 0.214, respectively. It indicates that as the extractable value of bundle MEV activities increases, transaction waiting times in blocks correspondingly increase~\cite{myers2004spearman}. Hence, the results provide further evidence of bundle MEV activities on delaying transaction waiting times.

\noindent\textbf{Summary}: \textit{Bundle MEV activities endanger blockchain users' economic security by delaying users' transactions in later blocks.}

\subsection{Other feasible applications}
\label{sec_other_applications}

\noindent
$\bullet$ Our approach can automatically uncover the evolving strategies of arbitrageurs (e.g.,~\cite{bartoletti2021maximizing}) for extracting MEV from new DeFi applications in practice. 
As a result, our approach uncovered 17 new MEV strategies on five kinds of popular DeFi applications (i.e., AMM, Lending, NFT, Airdrop, and Rebase Token) in \S\ref{sec_rq3evaluation}.
Compared to us, existing studies~\cite{zhou2021just,zhou2021high,babel2021clockwork,bartoletti2021maximizing} only manually design strategies for extracting MEV from a few DeFi applications that are well-studied (e.g., AMM). The new strategies discovered by our approach can motivate researchers to further explore the design space of MEV strategies. We use~\cite{babel2021clockwork} as an example to illustrate how our uncovered MEV strategies can facilitate relevant studies~\cite{zhou2021just,zhou2021high,babel2021clockwork,bartoletti2021maximizing} to design new strategies to extract MEV from these applications. Specifically, in Rebase Backrun Arbitrage (RBA in \S\ref{sec_rq3evaluation}), we found that an arbitrageur can backrun a Rebase token to steal assets in AMMs with zero cost. Please note that~\cite{babel2021clockwork} explores possible MEV activities on AMMs by only manually modeling AMMs. Our findings on RBA can help~\cite{babel2021clockwork} design new MEV strategies for AMMs (e.g., RBA) by including the modeling of the Rebase token.

\noindent
$\bullet$
Considering the fast-growing amount of bundles (e.g., there are more than 6,641,481 bundles and 26,740,394 transactions in bundles until Dec. 2022), our approach can be used to detect and discover MEV activities in bundles continuously. 
As a result,
by automatically analyzing 6,641,481 bundles, our approach detected 2,112,344 bundles containing known MEV activities, and 3,182,363 bundles only containing new MEV activities (\S\ref{sec_hunter}). 
Compared to us, existing work~\cite{Daian2020flash,qin2021quantifying,liu2022empirical,ferreira2021frontrunner} on quantifying MEV activities involves lots of manual efforts.
Our approach can benefit relevant studies~\cite{Daian2020flash,qin2021quantifying,liu2022empirical,ferreira2021frontrunner} through using our approach and its results.
For instance,~\cite{qin2021quantifying} can leverage our approach to detect MEV activities, and then leverage its results to conduct an in-depth study (e.g.,~\cite{qin2021quantifying} investigated arbitrageurs' strategies in  MEV activities).

\noindent
$\bullet$ 
Our approach can recognize stealthy attacks. Stealthy attacks are launched by 
attackers via exploiting bundles. Without bundles, attackers have to broadcast their transactions in the P2P network, and their transactions may also be attacked by other attackers. %
Our approach recognized three stealthy attacks in \mevformula{D$_{\mevsmallformula{Bundle}}$} (cf. details in Appendix H) through \cluster{} when we leveraged \cluster{} to discover DeFi MEV activities (\S\ref{sec_rq3evaluation}).
Although Zhou et al.~\cite{zhou2022sok} reported stealthy attacks, our approach distinguishes them in three points: i) different from Zhou et al.~\cite{zhou2022sok} that collected attacks from literature and confirmed stealthy ones manually, our approach can automatically recognize stealthy attacks through \cluster{} by recognizing outliers; ii) Given the fast-growing amount of bundles, the emerging new DeFi applications, and the evolving MEV strategies, our approach empowers the continuous detection and recognition of stealthy attacks, whereas Zhou et al.~\cite{zhou2022sok} only collected stealthy attacks from literature; iii) Our approach categorizes stealthy attacks by clustering them in \cluster{}, whereas Zhou et al.~\cite{zhou2022sok} manually categorized two types of stealthy attacks.

\section{Threats to Validity}
\label{sec_discussion}

Due to the lack of ground-truth dataset, we manually analyze DeFi actions identified by \lifter{} and baseline techniques (e.g., Etherscan and DeFiRanger).
Since manual inspection is labor-intensive, we did not check whether there is any DeFi action missed by all techniques, and thus the result of false negatives might be affected. In the future, we will 
involve more efforts to 
inspect all 6,641,481 bundles to detect FNs
missed by all techniques.

It raises threats to validity that we do not analyze all relays' bundles in blockchains, since different relays on a chain and relays on different chains have different strategies for relaying bundles (e.g., whether to follow censorship~\cite{mevrelay2021list}).
However, these strategies do not change how bundle arbitrageurs perform MEV activities (e.g., manipulating transactions’ positions). Hence, our approach can be generalized to bundle MEV studies in the wider ecosystem.
Moreover, we have analyzed multiple relays’ bundles.
It is worth noting that, for relays disclosing bundles relayed by them, Flashbots will collect their bundles and list
them
in Flashbots' web API~\cite{flashbots2021api}.
Hence, the bundles 
collected in~\S \ref{sec_datacollection} also contain
bundles relayed by other disclosed bundle relays, e.g., Eden~\cite{eden2021bundles}.  
In future work, we will investigate relays that do not disclose their relayed bundles.

The completeness for representing DeFi actions in bundles into the low-dimensional feature space (\S\ref{sec_actcluster}) and labeling bundles with MEV activities in the feature space (\S\ref{sec_actcluster}) cannot be provably guaranteed due to the lack of ground truth. 
Thus, MEV activities involving other DeFi actions (that are not in \textbf{A1}-\textbf{10} in \S\ref{sec_DeFiAppAct}) can be missed as false negatives.
Although our ten manually selected DeFi actions (which are heavily involved in MEV activities) are by no means complete, our approach can be easily extended to discover more new DeFi MEV activities by including more kinds of DeFi actions. 
In future work, we will inspect more kinds of DeFi actions.

The delayed transactions can result from various factors such as MEV activities, P2P network congestion~\cite{liu2022empirical}, and gas fee volatility of transactions~\cite{liu2022empirical}. Our Granger causality test (\S\ref{sec_network_security}) determines that bundle MEV activities can contribute to the delay of users' transactions.
In future work, we will explore to what extent these factors contribute to increase delays of users' transaction.

While collecting DeFi actions from Etherscan (\S\ref{sec_rq2}), we have taken ethical considerations by limiting our collection of DeFi actions to a slow pace (i.e., querying one page per ten seconds) and manually solving the reCAPTCHA human authentication.
However, our collection of DeFi actions from Etherscan still goes against Etherscan's terms~\cite{etherscan2015etherscan}, and it potentially raises questions about the ethicality of the collection process for DeFi actions from Etherscan.

\section{Related work}
\label{sec_relatedwork}

We introduced four categories of closely-related work. 

\noindent\textbf{DeFi action identification.}
Majority of existing studies~\cite{wang2021cyclic, qin2021quantifying, qin2021empirical, wang2022speculative, piet2022extracting, explore2021bundles,weintraub2022flash}
only focus on a few DeFi applications and could not cover other DeFi applications. We compared them in~\S \ref{sec_intro} and~\S \ref{sec_preparation}.
Etherscan~\cite{etherscan2015etherscan} identifies 7 kinds of DeFi actions, and DeFiRanger~\cite{wu2021defiranger} automatically recognizes DeFi actions.
However, both of them suffer inaccurate results, and 
our approach outperforms them (\S \ref{sec_rq2}). %

\noindent\textbf{Design on extracting MEV.} 
Eskandari et al.~\cite{Eskandari2020sok} introduce the front-running taxonomy.
Zhou et al.~\cite{zhou2021just} generate profitable MEV activities by interacting with AMMs.
Zhou et al.~\cite{zhou2021high} formalize Sandwich Attacks %
with crafted Swap actions on AMMs.
Several studies (e.g.,~\cite{bartoletti2021maximizing, qin2021attacking}) model specific kinds of MEV activities, and determine optimal parameters to maximize the revenue of extracting MEV.
None of them can be used to conduct a systematic study on DeFi MEV activities, because they cannot recognize DeFi MEV activities with unknown patterns of DeFi actions.

\noindent\textbf{MEV evaluation.}
Existing studies only quantify known MEV activities,
and cannot discover unknown MEV activities.
Torres et al.~\cite{ferreira2021frontrunner} 
measure three types of front-running.
Daian et al.~\cite{Daian2020flash} 
evaluate the front-running under the gas price auction. 
Qin et al.~\cite{qin2021quantifying} quantify 
five kinds of MEV
activities.
For known MEV activities, several studies~\cite{qin2021empirical, wang2021cyclic, piet2022extracting, weintraub2022flash, wang2022impact,lyu2022empirical} evaluate their impact, users’ perceptions of them, and their prevalence in private transactions.

\noindent\textbf{MEV mitigation.}
Researchers propose countermeasures to mitigate threats caused by known MEV activities, and our insights from new DeFi MEV activities can contribute to them.
One solution is to guarantee the transaction order fairness (e.g.,
~\cite{kelkar2020order, kelkar2021themis})
so that validators/miners and traders cannot modify transaction positions to extract MEV. 
Other studies propose new blockchain platforms or applications to prevent front-running~\cite{breidenbach2018enter, LibSubmarine2021LibSubmarine, zhou2021a2mm, heimbach2022eliminating}. 
Furthermore,
several studies~\cite{baum2021sok,heimbach2022sok,yang2022sok} systematize countermeasures against the front-running, transaction reordering manipulation, and MEV, and discuss the corresponding attacks and open challenges. 

\section{conclusion}
\label{conclusion}

We conduct the first systematic study on DeFi MEV activities in Flashbots bundle by developing \lifter{}, a novel automated tool for accurately identifying DeFi actions in transactions, and \cluster{}, a new approach that leverages iterative clustering to facilitate the discovery of DeFi MEV activities. 
Our experimental results show that \lifter{} achieves nearly 100\% accuracy in identifying DeFi actions, significantly outperforming existing techniques. With the help of \cluster{}, we discover 17 new kinds of DeFi MEV activities, which occur in 53.12\% of bundles but have not been reported.
Moreover, we demonstrate that \lifter{} and \cluster{} are very useful in MEV studies by six applications.

\section*{Acknowledgements}
The authors thank the anonymous reviewers for their constructive comments.
This work is partly supported by Hong Kong RGC Projects (No. PolyU15219319, PolyU15222320, PolyU15224121), and National Natural Science Foundation under Grant No. 62202405.

\bibliographystyle{ACM-Reference-Format}
\balance
\bibliography{reference}

\input{appendix_to_remove}

\end{document}

%% file: solidity-highlighting.tex
\definecolor{verylightgray}{rgb}{.97,.97,.97}

\lstdefinelanguage{Solidity}{
	keywords=[1]{emit, anonymous, assembly, assert, balance, break, call, callcode, case, catch, class, constant, continue, contract, debugger, default, delegatecall, delete, do, else, event, export, external, false, finally, for, function, gas, if, implements, import, in, indexed, instanceof, interface, internal, is, length, library, log0, log1, log2, log3, log4, memory, modifier, new, payable, pragma, private, protected, public, pure, push, require, return, returns, revert, selfdestruct, send, storage, struct, suicide, super, switch, then, this, throw, transfer, true, try, typeof, using, value, view, while, with, addmod, ecrecover, keccak256, mulmod, ripemd160, sha256, sha3}, 
	keywordstyle=[1]\color{blue}\bfseries,
	keywords=[2]{address, bool, byte, bytes, bytes1, bytes2, bytes3, bytes4, bytes5, bytes6, bytes7, bytes8, bytes9, bytes10, bytes11, bytes12, bytes13, bytes14, bytes15, bytes16, bytes17, bytes18, bytes19, bytes20, bytes21, bytes22, bytes23, bytes24, bytes25, bytes26, bytes27, bytes28, bytes29, bytes30, bytes31, bytes32, enum, int, int8, int16, int24, int32, int40, int48, int56, int64, int72, int80, int88, int96, int104, int112, int120, int128, int136, int144, int152, int160, int168, int176, int184, int192, int200, int208, int216, int224, int232, int240, int248, int256, mapping, string, uint, uint8, uint16, uint24, uint32, uint40, uint48, uint56, uint64, uint72, uint80, uint88, uint96, uint104, uint112, uint120, uint128, uint136, uint144, uint152, uint160, uint168, uint176, uint184, uint192, uint200, uint208, uint216, uint224, uint232, uint240, uint248, uint256, var, void, ether, finney, szabo, wei, days, hours, minutes, seconds, weeks, years},	
	keywordstyle=[2]\color{teal}\bfseries, 
	keywords=[3]{block, blockhash, coinbase, difficulty, gaslimit, number, timestamp, msg, data, gas, sender, sig, value, now, tx, gasprice, origin},	
	keywordstyle=[3]\color{violet}\bfseries,
	keywords=[4]{onlyOwner},
	keywordstyle=[4]\color{red}\bfseries,
	identifierstyle=\color{black},
	sensitive=false,
	comment=[l]{//},
	morecomment=[s]{/*}{*/},
	commentstyle=\color{gray}\ttfamily,
	stringstyle=\color{red}\ttfamily,
	morestring=[b]',
	morestring=[b]"
}

%% file: appendix_to_remove.tex
\renewcommand{\thesection}{\Alph{section}}
\setcounter{section}{0}
\renewcommand\theHsection{appendix.\arabic{section}}

\section{Example of event extraction}
\label{sec_appendix_eventextraction}
Fig.~\ref{fig_uniswapexample_a} shows Uniswap's~\cite{Adams2020UniswapVC} descriptions of \texttt{\small Swap} event.
We confirm it corresponds to a Swap action by its descriptions in Line 8.
Fig.~\ref{fig_smoothyexample_b} shows code snippets and comments of another \texttt{\small Swap} event and \texttt{\small swap} function from Smoothy (\url{https://smoothy.finance/}). We confirm it corresponds to a Swap action by comments in Line 2 and the two functions which will trigger asset transfers in Line 5 and 6.

\begin{figure}[h]
\small
  \begin{subfigure}{\linewidth}
    \begin{lstlisting}[mathescape=true,language=Solidity, frame=none, basicstyle=\linespread{0.6} \fontsize{6}{9}\ttfamily]
event Swap {
    address indexed sender,
    uint amount0In,
    uint amount1In,
    uint amount0Out,
    uint amount1Out,
    address indexed to};
// Emitted each time a swap occurs via swap function.
    \end{lstlisting}
    \caption{descriptions of \texttt{\small Swap} event in Uniswap's document}
    \label{fig_uniswapexample_a}
  \end{subfigure}

  \begin{subfigure}{\linewidth}
    \begin{lstlisting}[mathescape=true,language=Solidity, frame=none, basicstyle=\linespread{0.6} \fontsize{6}{9}\ttfamily]
event Swap{...};
//* @dev Swap a token to another.
function swap(...){
    ...
    _transferIn(infoIn, bTokenInAmount);
    _transferOut(infoOut, bTokenOutAmount, adminFee);
    emit Swap(...);}
    \end{lstlisting}
    \caption{code snippets and comments of \texttt{\small Swap} event in Smoothy's codes}
    \label{fig_smoothyexample_b}
  \end{subfigure}
  \caption{Event information of Uniswap and Smoothy.}
  \label{fig_manualexample}
\end{figure}

\section{Recognize asset transfers}
\label{sec_appendix_recognize_asset_transfer}

\noindent$\bullet$
\textbf{Token transfer.} 
In a token transfer \mevformula{Asset}$_{\mevsmallformula{C}}$\mevformula{.Transfer(From, To, Value)}, \mevformula{From} sends \mevformula{Value} amounts of \mevformula{Asset}$_{\mevsmallformula{C}}$ to \mevformula{To}.
Hence, \mevformula{c$_1$}, denoted as \mevformula{C.Event(Transfer(\\From,To,Value))}, checks whether a token transfer occurs when a \texttt{\small Transfer} event is emitted by \mevformula{C} with the parameters of \mevformula{From}, \mevformula{To}, and \mevformula{Value}.
A token transfer should satisfy three requirements in \mevformula{c$_2$}: i)
\mevformula{From} is not the zero address and \mevformula{C}'s address (i.e., \mevformula{From} $\not\in$ \mevformula{(0x00...00, C)}), and \mevformula{To} is also not the zero address and \mevformula{C}'s address (i.e., \mevformula{To} $\not\in$ \mevformula{(0x00...00, C)}). This requirement is based on the widely used templates for ERC20 and ERC721 (e.g., OpenZeppelin~\cite{OpenZeppelinERC20721template} and chiru-labs~\cite{chirulabsERC721template}). In their templates, the zero address and the address of \mevformula{C} are used for token minting and burning.
ii) the amount of transferred token \mevformula{Value} is non-zero (i.e., \mevformula{Value} $\neq$ \mevformula{0}), and iii) \mevformula{From} and \mevformula{To} are different addresses (i.e., \mevformula{From} $\neq$ \mevformula{To}).
Note that there is no actual asset transfer between \mevformula{From} and \mevformula{To} if any of the last two requirements are violated.

\noindent$\bullet$
\textbf{ERC721 token minting/burning.}
In an ERC721 token minting (resp. burning) \mevformula{Asset}$^{721}_{\mevsmallformula{C}}$\mevformula{.Minting(From, To, Value)} (resp. \mevformula{Asset}$^{721}_{\mevsmallformula{C}}$\mevformula{.Burning(From, To, Value)}), the ERC721 token contract \mevformula{C} mints (resp. burns) an NFT with the tokenId \mevformula{Value}.
Hence, \mevformula{c$_1$}, denoted as \mevformula{C.Event(Transfer(From,To,Value))}, checks whether an ERC721 token minting (resp. burning) occurs when \mevformula{C} emits a \texttt{\small Transfer} event with the parameters of \mevformula{From}, \mevformula{To}, and \mevformula{Value}.
An ERC721 token minting (resp. burning) should satisfy two requirements in \mevformula{c$_2$}: i) 
\mevformula{From} (resp. \mevformula{To}) is the zero address or \mevformula{C}'s address (i.e., \mevformula{From} (resp. \mevformula{To}) $\in$ \mevformula{(0x00...00, C)}), and \mevformula{To} (resp. \mevformula{From}) is not the zero address and \mevformula{C}'s address (i.e., \mevformula{To} (resp. \mevformula{From}) $\not\in$ \mevformula{(0x00...00, C)}).
This requirement is based on the widely used templates for ERC721 (e.g., OpenZeppelin~\cite{OpenZeppelinERC20721template} and chiru-labs~\cite{chirulabsERC721template}). In their templates, the zero address and \mevformula{C}'s address are used for ERC721 token minting and burning.
ii) \mevformula{C} implements standard functions defined in ERC721 (i.e., \mevformula{C} $\models$ \mevformula{ERC721 standard}), and thus,
\mevformula{Asset$_{\mevsmallformula{C}}$} is an ERC721 asset.

\section{Alternative approach of \lifter{}.}
To evaluate the effectiveness of collected events in $\mathbb{M}$, 
we created two variants of \lifter{}.
\lifter{}$_{a1}$ replaces $\mathbb{M}$ (\S \ref{sec_preparation}) with other information collected more automatically, i.e., contract addresses of DeFi applications.
Besides, in \textbf{S-1}, \lifter{}$_{a1}$ recognizes asset transfers involved in DeFi actions, if the contract addresses of DeFi applications receive or send assets in the asset transfers.
Then \lifter{}$_{a1}$ identifies DeFi actions in \textbf{S-2}.
To obtain the contract addresses, we queried the APIs of graph, which provides blockchain data to developers~\cite{thegraph}.
\lifter{}$_{a2}$ ignores $\mathbb{M}$, only recognizes all asset transfers in transactions in \textbf{S-1}, and identifies DeFi actions in \textbf{S-2}.

We compare \lifter{} with its two variants by 500 transactions.
It shows that \lifter{}$_{a1}$ reports 7 false actions, because asset transfers can simultaneously satisfy asset transfer patterns of different kinds of actions, 
and a DeFi contract can perform different kinds of actions. 
By only using address information, \lifter{}$_{a1}$ cannot distinguish DeFi actions performed by the contract.
\lifter{}$_{a2}$ reports 87 false actions performed by non-DeFi contracts due to wrongly pairing asset transfers, because \lifter{}$_{a2}$ identifies DeFi actions only according to asset transfer patterns.
By contrast, \lifter{} correctly identifies all DeFi actions with $\mathbb{M}$.

\section{Frequency of events in $\mathbb{M}$}
\label{sec_eventfrequency}
\begin{figure}[!b]
	\centering
	\includegraphics[width=0.99\linewidth]{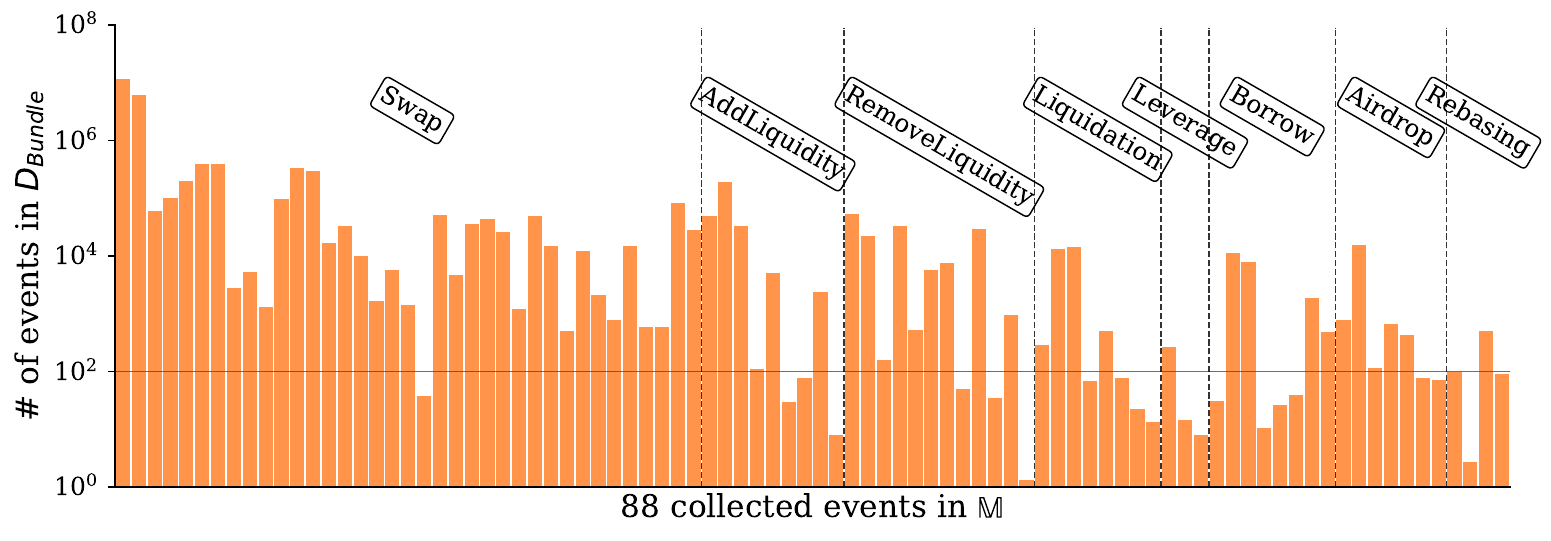}
	\caption{Frequency of events in $\mathbb{M}$}
	\label{fig_eventfrequency}
\end{figure}

To ensure comprehensive and reliable results, we further assess whether all 88 events in $\mathbb{M}$ (\S\ref{sec_preparation}) occur in \mevformula{D$_{\mevsmallformula{Bundle}}$}.
Fig.~\ref{fig_eventfrequency} displays the frequency of each event in $\mathbb{M}$ (\S\ref{sec_preparation}) that occurs in \mevformula{D$_{\mevsmallformula{Bundle}}$}.
We find that all 88 events are covered, with 64 of them (72.7\%) occurring at least 100 times in \mevformula{D$_{\mevsmallformula{Bundle}}$}. Additionally, the results indicate that methods (e.g.,~\cite{wang2021cyclic, qin2021quantifying, qin2021empirical, wang2022speculative, piet2022extracting, explore2021bundles,weintraub2022flash} in Table~\ref{table_related_technique_com}) relying on several specific events to identify DeFi actions will miss reporting a significant number of DeFi actions.

\section{New MEV activities in Bundles}
\label{appendix_10mev}

We discover 17 kinds of new DeFi MEV activities in bundles, which are summarized in Table~\ref{table_11mevactsdest}.
For a more thorough understanding of them, we refer readers to the previous helpful studies~\cite{ferreira2021frontrunner,qin2021empirical,wang2021cyclic} which provide basic descriptions for the three known MEV activities, i.e., Sandwich Attack, Cyclic Arbitrage, and Liquidation.

\subsection{Multi-layered Burger Arbitrage (MBA)}
It involves more than three transactions in a bundle. The first and last transactions are emitted by the arbitrageur as \mevformula{A$_1$} and \mevformula{A$_2$}, and all the other transactions are emitted by other traders as \mevformula{V$_1$,...,V$_n$}, where \mevformula{n} $>$ 1. 
All \mevformula{V$_i$} and \mevformula{A$_1$} aim to trade \mevformula{X} asset for \mevformula{Y} asset in the same AMM, and \mevformula{A$_2$} aims to trade \mevformula{Y} for \mevformula{X} with the same AMM.
The Multi-layered Burger Arbitrage is similar to Sandwich Attack~\cite{qin2021quantifying}, except that there are more than one transaction in the middle of \mevformula{A$_1$} and \mevformula{A$_2$} transactions. All \mevformula{V$_i$} are used to pull up the price of \mevformula{Y} in the AMM, and further improve the arbitrageur's revenue.
For example, an arbitrageur performed the Multi-layered Burger Arbitrage in the first bundle of the 
12,753,463 block~\cite{mbaexample2021mbaexample}. In the bundle, there are 4 victim transactions. The four victim transactions all trade \texttt{\small ETH} for \texttt{\small F9} with AMM \texttt{\small 0x459e}. The arbitrageur earns 1.16 \texttt{\small ETH} as profits by the \mevformula{A$_1$} and \mevformula{A$_2$} in the bundle.

\begin{figure}[b!]
	\centering
	\includegraphics[width=0.4\textwidth]{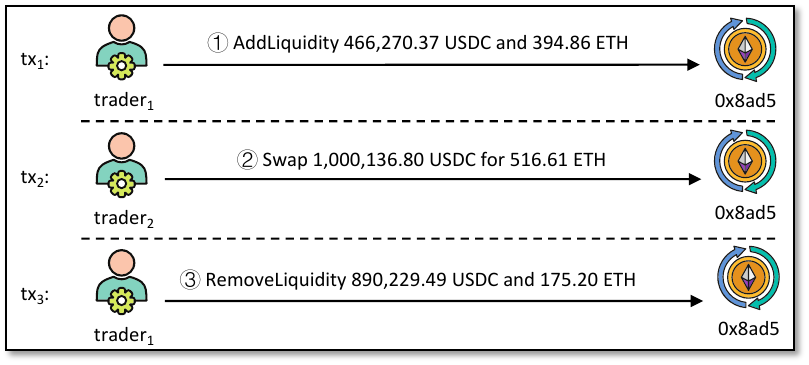}
	\caption{ An example of Liquidity Sandwich Arbitrage}
	\label{fig_lsaexample}
\end{figure}

\subsection{Liquidity Sandwich Arbitrage (LSA)}
\label{sec_appendix_lsa}
It involves three transactions.
The first and third transactions are signed by the trader$_1$ as \mevformula{A$_1$} and \mevformula{A$_2$}, and the second transaction is signed by trader$_2$ as \mevformula{V}. The \mevformula{V} aims to trade \mevformula{X} asset for \mevformula{Y} asset in an AMM, \mevformula{A$_1$} and \mevformula{A$_2$} aim to supply and withdraw \mevformula{X} and \mevformula{Y} assets with the same AMM. Different from Sandwich Attack~\cite{qin2021quantifying}, Liquidity Sandwich Arbitrage does not aim to pull up the price of the traded assets like \mevformula{Y} when \mevformula{V} executes, but the trader$_1$ aims to be the liquidity provider to earn the exchange fee for the swapping in \mevformula{V}.
In fact, \mevformula{V} can set the slippage protection parameter~\cite{heimbach2022eliminating} to require the minimum amount of received assets, if trader$_1$ conducts the Sandwich Attack, the slippage protection parameter can trigger and the execution of \mevformula{V} will revert.
Hence the Sandwich Attack gains no profits.
For example, Fig.~\ref{fig_lsaexample} shows the three transactions in the first bundle of the 12,702,238 block. In the first and third transactions, the trader$_1$ supplies and withdraws \texttt{\small USDC} and \texttt{\small ETH} with the AMM \texttt{\small 0x8ad5}, respectively, and in the second transaction, trader$_2$ trades \texttt{\small USDC} for \texttt{\small ETH} in the AMM \texttt{0x8ad5}.
According to the price in Etherscan~\cite{etherscan2015etherscan} of \texttt{\small ETH} and \texttt{\small USDC} on the day of mining the 12,702,238 block, trader$_1$ earns profits of 18637.5 USD.

\noindent
\textbf{\textit{Insight.}}
The observation from LSA yields the security insights for MEV countermeasures implemented in the contracts (e.g., slippage protection~\cite{heimbach2022eliminating}, atomic routing~\cite{zhou2021a2mm}, and optimal slippage setting~\cite{heimbach2022eliminating}]). More precisely, these MEV countermeasures rely on the parameters in contracts to defend against MEV (e.g., failing the transactions where parameters are triggered). However, bundle arbitrageurs can still maximize their revenue without triggering the MEV protection mechanisms implemented in the contracts, because bundle arbitrageurs can manipulate the order of transactions in their bundles to make their arbitrage transactions and victim transactions execute in bundle arbitrageurs' expected order.

\subsection{Backrun Cyclic Arbitrage (BCA)}
It involves two transactions.
In the first one, trader$_1$ performs Swap, AddLiquidity, or RemoveLiquidity actions, the actions trigger the unbalanced prices among the AMMs~\cite{qin2021quantifying}. 
In the second one, trader$_2$ backruns the former transaction to gain profit among the AMMs by Cyclic Arbitrage. %

\begin{figure}
	\centering
	\includegraphics[width=0.4\textwidth]{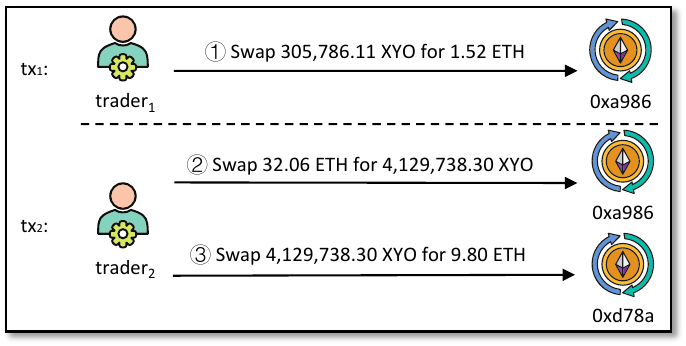}
	\caption{An example of Failed Arbitrage}
	\label{fig_faexample}
\end{figure}

\subsection{Failed Arbitrage (FA)}
For the Failed Arbitrage, the arbitrageur aims to obtain profits by performing Sandwich Attack or Cyclic Arbitrage in a bundle, and suffers the financial loss. 
For example, Fig.~\ref{fig_faexample} shows the two transactions in the second bundle of the 12,516,458 block. In the first transaction, trader$_1$ trades \texttt{\small XYO} for \texttt{\small ETH} in AMM \texttt{\small 0xa986}. In the second transaction, trader$_2$ aims to conduct the Cyclic Arbitrage to backrun the first transaction in the AMM \texttt{0xa986}. Unfortunately, the trader$_2$ suffers the financial loss of 22.26 (32.06-9.8) Ether due to trading \texttt{\small XYO} for \texttt{\small ETH} in the AMM \texttt{\small 0xd78a}.

\subsection{Hybrid Arbitrage (HA)}
For the Hybrid Arbitrage, 
there are at least two kinds of MEV activities of the three known MEV activities (i.e., Sandwich Attack, Cyclic Arbitrage, and Liquidation) in a bundle. Besides, the occurrence of transactions of the MEV activities is crossed.
Hybrid Arbitrage activities are the cases when arbitrageurs perform multiple kinds of MEV activities in a bundle. Besides, to minimize the transaction fee cost, arbitrageurs merge multiple kinds of MEV activities into a single transaction. For example, in a transaction, an arbitrageur first performs a Liquidation activity to receive the collateral assets, and then uses received assets to perform a Cyclic Arbitrage activity.

\subsection{Swap Backrun Arbitrage (SBA)}
It involves two transactions. The former executes a Swap action to exchange \mevformula{X} asset for \mevformula{Y} asset in an AMM which pulls up \mevformula{Y}'s price, and the latter backruns the former transaction by exchanging \mevformula{Y} asset for \mevformula{X} asset in the same AMM to sell \mevformula{Y} at the higher price than without the former transaction.

\begin{figure}[t!]
	\centering
	\includegraphics[width=0.4\textwidth]{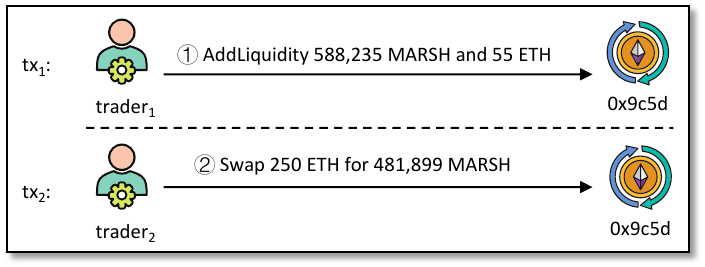}
	\caption{ An example of Liquidity Backrun Arbitrage}
	\label{fig_lbaexample}
\end{figure}

\subsection{Liquidity Backrun Arbitrage (LBA)}
It involves two transactions.
The former executes an AddLiquidity/RemoveLiquidity action on an AMM which causes the unbalanced prices between AMMs, and the latter executes a Swap action to trade the corresponding assets on the same AMM to obtain profits from the price differences.
For example, Fig.~\ref{fig_lbaexample} shows the two transactions in the first bundle of the 12,141,301 block. In the former transaction, trader$_1$ supplies \texttt{\small MARSH} and \texttt{\small ETH} to the AMM \texttt{\small 0x9c5d}. In the latter transaction, trader$_2$ performs a Swap to trade \texttt{\small ETH} for \texttt{\small MARSH} with the same AMM.
According to the price in Etherscan~\cite{etherscan2015etherscan} of \texttt{\small MARSH} and \texttt{\small ETH} on the day of mining the 12,141,301 block, in the latter transaction, the trader$_2$ trades \texttt{\small ETH} for \texttt{\small MARSH} which are worth 738,060 and 3,504,910.65 USD, respectively.

\subsection{Liquidity-swap Trade (LT)}
\begin{figure}
	\centering
	\includegraphics[width=0.4\textwidth]{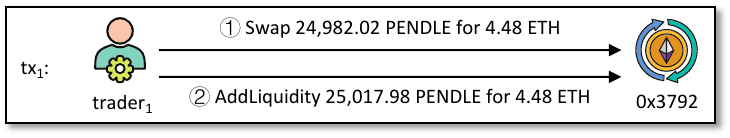}
	\caption{An example of Liquidity-swap Trade}
	\label{fig_ltexample}
\end{figure}

For the Liquidity-swap Trade, there exists a transaction that a trader both trades assets on AMMs and performs the AddLiquidity or RemoveLiquidity actions on AMMs. There are two cases for the Liquidity-swap Trade, i) the trader trades assets and supplies the traded assets into an AMM at the expected price, ii) the trader withdraws assets from an AMM and trades the returned assets. 
For example, Fig.~\ref{fig_ltexample} shows the transaction in the second bundle of the 13,521,679 block, the trader trades \texttt{\small PENDLE} for \texttt{\small ETH} and supplies \texttt{\small PENDLE} and the traded \texttt{\small ETH} to the AMM \texttt{\small 0x3792}. By the transaction, the trader becomes the liquidity provider at the expected price of assets in the AMM \texttt{\small 0x3792}.

\subsection{Partial Cyclic Arbitrage (PCA)}
For the Partial Cyclic Arbitrage, there exists a transaction that performs multiple Swap actions among AMMs. Part of the Swap actions can fit into a single cycle one by one in the transaction.
The Partial Cyclic Arbitrage distinguishes the Cyclic Arbitrage,
because Cyclic Arbitrage only considers the transactions in which all the Swap actions fit into a single cycle.

\subsection{Non-cyclic Swap Trade (NST)}
\begin{figure}
	\centering
	\includegraphics[width=0.35\textwidth]{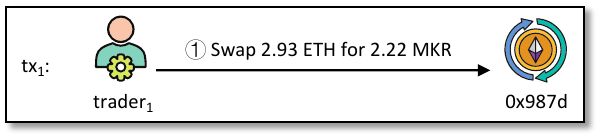}
	\caption{An example of Non-cyclic Swap Trade}
	\label{fig_nstexample}
\end{figure}

For the Non-cyclic Swap Trade,
the transactions in bundles only perform the Swap actions among AMMs. Besides, there are no known MEV activities, i.e., Sandwich Attack, Cyclic Arbitrage, and Liquidation, no the other 9 kinds of DeFi actions, and no other 16 kinds of new MEV activities. The trader who performs the Non-cyclic Swap Trade in a bundle aims to trade on the AMMs at the expected price.
For example, Fig.~\ref{fig_nstexample} shows the transaction in the first bundle of the 12,244,578 block. The trader$_1$ only trades \texttt{\small ETH} for \texttt{\small MKR} with the AMM \texttt{\small 0x987d}. According to the price in Etherscan~\cite{etherscan2015etherscan} of \texttt{\small ETH} and \texttt{\small MKR} on the day of mining the 12,244,578 block, the trader$_1$ trades \texttt{\small ETH} for \texttt{MKR} which are worth 7,356.07 and 8,375.68 USD, respectively. Hence, trader$_1$ obtains profits as 1,019.61 USD.

\subsection{Bulk NFT-Minting (BN)} It only involves transactions that mint NFTs. There are two cases for the Bulk NFT-Minting, i) the NFT contract mints multiple NFTs in a single transaction, ii) the NFT contract mints multiple NFTs among multiple transactions.
For example, there is one transaction in the second bundle of the 13,336,591 block, and the \texttt{\small VIXEN} NFT contract mints 20 NFTs. %

\subsection{NFT Reforging (NR)} It involves one transaction, where the NFT contract first burns one NFT and then remints the NFT with the same tokenId and the specific asset represented by the NFT is updated. For example, the tenth bundle at 14,579,991 block contains one transaction, where the resolved address of the \texttt{\small ENS} NFT with the name \texttt{\small blockchainpolice.eth} is updated to address \texttt{\small 0x6669}.

\subsection{Airdrop Claiming (AC)}
For the Airdrop Claiming, a trader only claims airdrop rewards in transactions of a bundle.

\subsection{NFT-Minting-swap Trade (NT)}
It involves a transaction where a trader first receives a minted NFT, then the trader conducts asset exchange to sell the minted NFT to an AMM.

\subsection{Loan-powered Arbitrage (LA)}
It involves a transaction where an arbitrageur first loans assets from Lending under the over/under-collateral deposit, then uses the loaned assets to conduct MEV activities, e.g., Cyclic Arbitrage.

\subsection{Airdrop-swap Trade (AT)} It involves a transaction, where a trader both claims the airdrop rewards and sells the received assets to an AMM by swapping.

\section{Bundle formatting}
\label{sec_appendix_bundleformatting}

\noindent
$\bullet$
\textbf{Action block.}
An action block consists of parameters to characterize the corresponding type of action. 
It is a matrix where a column vector corresponds to either a parameter or a separator. %
Specifically,
the first column vector of each action block is the action header and it represents the action type with one-hot encoding.
The second column vector acts as a separator between the action header and other parameters, and all entries in it equal to $-$1.
We index all addresses involved in a transaction in chronological order and
represent them using their indices instead of the original addresses.
We also index different assets according to their popularity.
It is worth noting that the transferred amounts of assets are normalized in the range [$-$1, 1]
to avoid huge differences in parameter scale.

\noindent
$\bullet$
\textbf{Transaction block.}
It is comprised of meta information of a transaction and all actions in it.
The first (resp. second) column vector of a transaction block represents the sender (resp. recipient) of this transaction.
The last column vector records asset changes for different addresses after the transaction.
Between the second and last column vectors,
action blocks are sequentially concatenated and connected by separators.

\noindent
$\bullet$
\textbf{Bundle matrix.}
It is constructed by combining all transaction blocks within a bundle.
Transaction blocks are arranged in chronological order to reflect the temporal patterns related to DeFi actions in a bundle.
Successive transaction blocks are separated by two separators.
The height of bundle matrices equals the length of the column vector, whereas
the width of bundle matrices can be variable depending on the number of encapsulated transaction blocks.
To facilitate feature extraction and feature representation,
we fix the width of bundle matrices by specifying the maximum width.
The bundle matrix will be truncated if its width exceeds the maximum width, otherwise, it will be padded with $-$1 entries to fit the maximum width.

\begin{table}[b!]
\centering
\caption{Top 10 tokens involved in new MEV activities}
\resizebox{\linewidth}{!}{
\begin{tabular}{|l|ccccclllll|}
\hline
Token                                                          & WETH      & USDC    & USDT    & WBTC    & DAI    & BNT    & SHIB    & LINK    & APE    & SUSHI  \\ \hline
\rowcolor{black!20} \begin{tabular}[c]{@{}l@{}}\# MEV\\ activities\end{tabular} & 3,034,387 & 730,832 & 344,353 & 195,240 & 162,178 & 80,408 & 55,607 & 54,569 & 45,371 & 42,014 \\ \hline
\end{tabular}
}
\label{table_top10token}
\end{table}

\section{Prevalence of new MEV activities}
\label{sec_empiricalresult}

\begin{table*}[h!]
\small
\centering
\caption{Number of bundles for each kind of DeFi MEV activities}
\vspace{1pt}
\label{tab_kmeans_cnn2d}
\resizebox{\linewidth}{!}{
\begin{threeparttable}
\begin{tabular}{|l|cccccccccccccccccccc|}
\hline
MEV type & SA  & CA & LI  & SBA & LBA & LSA & MBA & LT & PCA & BCA & HA  & FA & NST & RBA & AT & BN & NR & AC & NT & LA \\ \hline
\rowcolor{black!20} \# bundles  & 813,188 & 1,334,207 & 14,263 & 162,375 & 5,045  & 12,830  & 3,654  & 5,578  & 65,670 & 46,771 & 70,413 & 46,784 & 3,160,094 & 54 & 128 & 16,327 & 218 & 2,388 & 562 & 2,470 \\ \hline
\end{tabular}
\begin{tablenotes}[flushleft]
{
\setlength{\itemindent}{-2.49997pt} \small
\item \footnotesize{SA: Sandwich Attack, CA: Cyclic Arbitrage, LI: Liquidation, SBA: Swap Backrun Arbitrage, LBA: Liquidity Backrun Arbitrage, LSA: Liquidity Sandwich Arbitrage, MBA: Multi-layered Burger Arbitrage, LT: Liquidity-swap Trade, PCA: Partial Cyclic Arbitrage, BCA: Backrun Cyclic Arbitrage, HA: Hybrid Arbitrage, FA: Failed Arbitrage, NST: Non-cyclic Swap Trade, RBA: Rebasing Backrun Arbitrage, AT: Airdrop-swap Trade, BN: Bulk NFT-Minting, NR: NFT Reforging, AC: Airdrop Claiming, NT: NFT-Minting-swap Trade, LA: Loan-powered Arbitrage.} 
}
\end{tablenotes}
\end{threeparttable}
}
\vspace{2pt}
\end{table*}

Table \ref{tab_kmeans_cnn2d} shows the number of bundles for each kind of MEV activities. The first row lists the type of DeFi MEV activities, and the second row lists the number of bundles containing corresponding DeFi MEV activities.

\noindent
$\bullet$
The number of bundles that contain different counts of new MEV types.
3,459,988 bundles only contain one kind of new MEV activities while 63,851 (resp. 3,816) bundles contain two (resp. three) kinds of new MEV activities.

\begin{figure}
    \centering
    \includegraphics[width=0.4\textwidth]{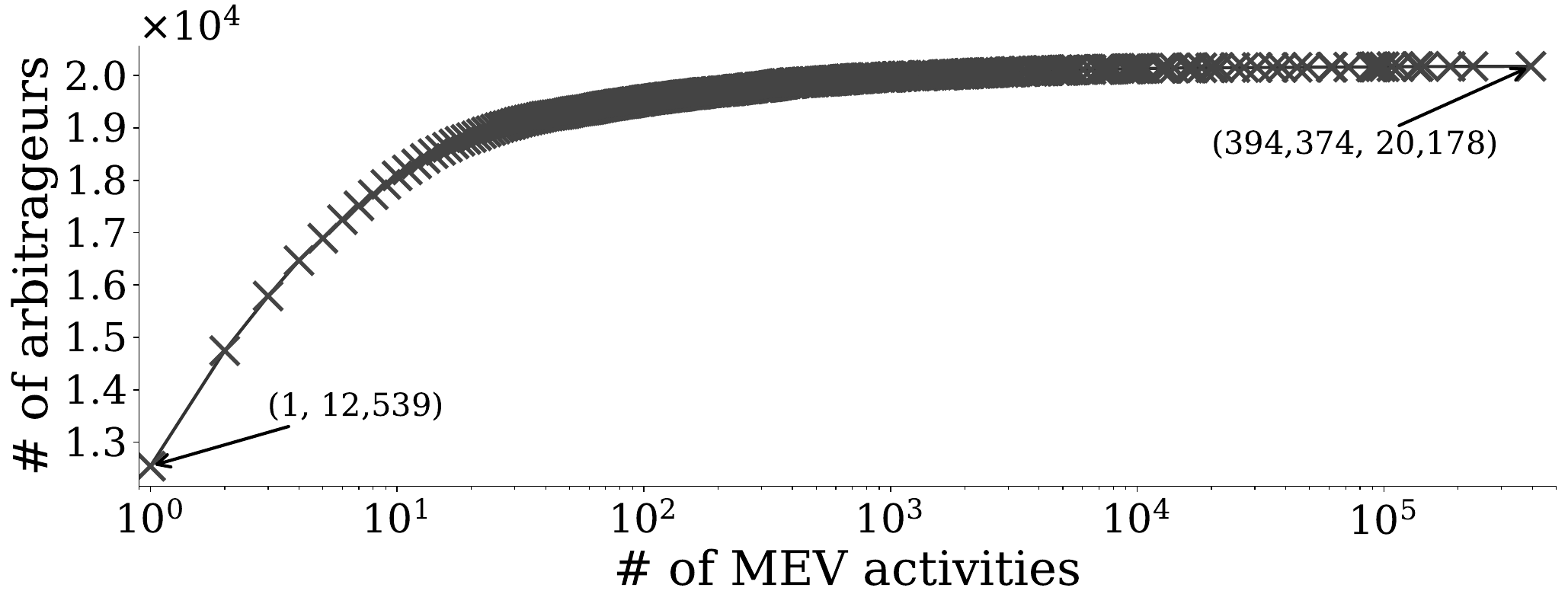}
    \caption{Number of contracts invoked directly by arbitrageurs to perform new DeFi MEV activities.} 
    \label{fig_personcount}
\end{figure}

\noindent
$\bullet$
The number of contracts that are directly invoked by the EOA account of arbitrageurs to perform new DeFi MEV activities.
In Fig.~\ref{fig_personcount}, each cross $(x, y)$ indicates
that $y$ contracts are invoked by arbitrageurs to conduct no more than $x$ MEV activities in all bundles. It shows
that 62.14\% (12,539/20,178) of contracts are involved in new DeFi MEV activities only once.
Table~\ref{table_top5contract} lists the top 5 contracts that are involved in new DeFi MEV
activities. The first, second, and third rows list the contract addresses, the
number of involved MEV activities, and the labels in Etherscan, respectively. The first contract is labeled as MEV Bot by Etherscan, and the first contract is directly invoked by arbitrageurs to conduct 11.18\% (394,374/3,527,655) of bundles containing new DeFi MEV activities. 
The other four contracts are labeled as AMM routers by Etherscan. It shows that the bundle arbitrageurs can calculate the parameters to perform MEV activities offline, and directly invoke the AMM routers to perform MEV activities.

\begin{figure}
    \centering
    \includegraphics[width=0.4\textwidth]{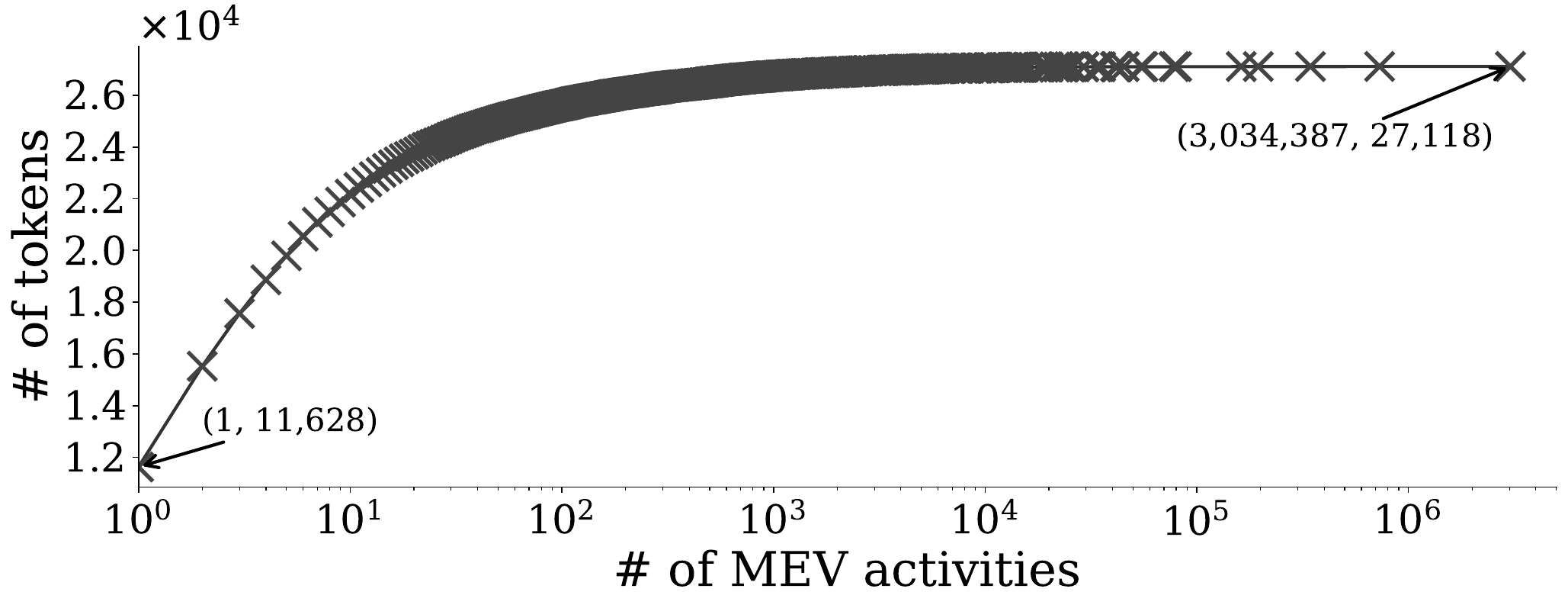}
    \caption{Number of tokens involved in new MEV activities.}   
    \label{fig_tokencount}
\end{figure}

\noindent
$\bullet$
The number of tokens used to perform new DeFi MEV activities by
arbitrageurs. In Fig.~\ref{fig_tokencount}, each cross $(x, y)$ indicates that $y$ tokens are involved in no more than $x$ MEV activities in all bundles. It shows that
42.88\% (11,628/27,118) of tokens are involved in new DeFi MEV activities only once.
Table~\ref{table_top10token} lists the top 10 tokens involved in new DeFi MEV
activities. The first and second rows list the tokens, and the number of involved MEV activities. It shows that \texttt{\small WETH} is involved in most of the new DeFi MEV activities (i.e., 86.02\%).

\begin{table}[t!]
\centering
\caption{Top 5 contracts involved in new MEV activities}
\resizebox{\linewidth}{!}{
\begin{tabular}{|l|ccccc|}
\hline
Contract &
  0xa57b...d6cf &
  0x7a25...488d &
  0x68b3...fc45 &
  0xe592...1564 &
  0xd9e1...8b9f \\ \hline
\rowcolor{black!20}\# MEV activities &
  394,374 &
  229,723 &
  186,143 &
  145,930 &
  137,818 \\ \hline
\rowcolor{black!5}Label &
  MEV Bot &
  \begin{tabular}[c]{@{}c@{}}Uniswap Router\end{tabular} &
  \begin{tabular}[c]{@{}c@{}}Uniswap Router\end{tabular} &
  \begin{tabular}[c]{@{}c@{}}Uniswap Router\end{tabular} &
  \begin{tabular}[c]{@{}c@{}}Sushiswap Router\end{tabular} \\ \hline
\end{tabular}
}
\label{table_top5contract}
\end{table}

\begin{figure}
    \centering
    \includegraphics[width=0.4\textwidth]{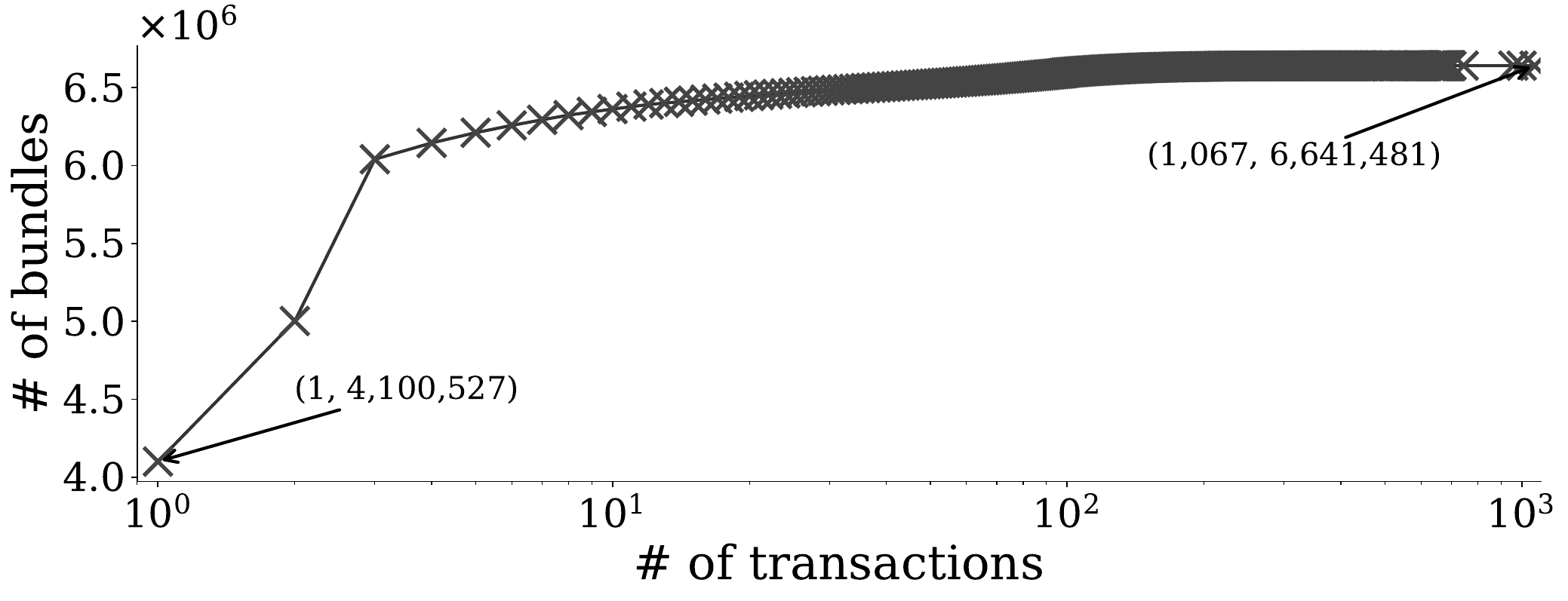}
    \caption{The number of transactions in each bundle.}   
    \label{fig_txcount}
\end{figure}

\noindent
$\bullet$
The number of transactions in each bundle. In Fig.~\ref{fig_txcount}, each cross $(x, y)$ indicates that $y$ bundles contain no more than $x$ transactions. 61.74\% bundles have only one transaction. Besides, the fourth bundle in the 13,143,462 block mined by F2Pool contains the most amount of transactions (i.e., 1,067). In the bundle, F2Pool distributes ETH to different EOA accounts for distributing mining revenues\footnote{\url{https://f2pool.io/mining/guides/how-to-mine-ethereumpow/}}. 

\begin{figure}[]
    \centering
    \includegraphics[width=0.4\textwidth]{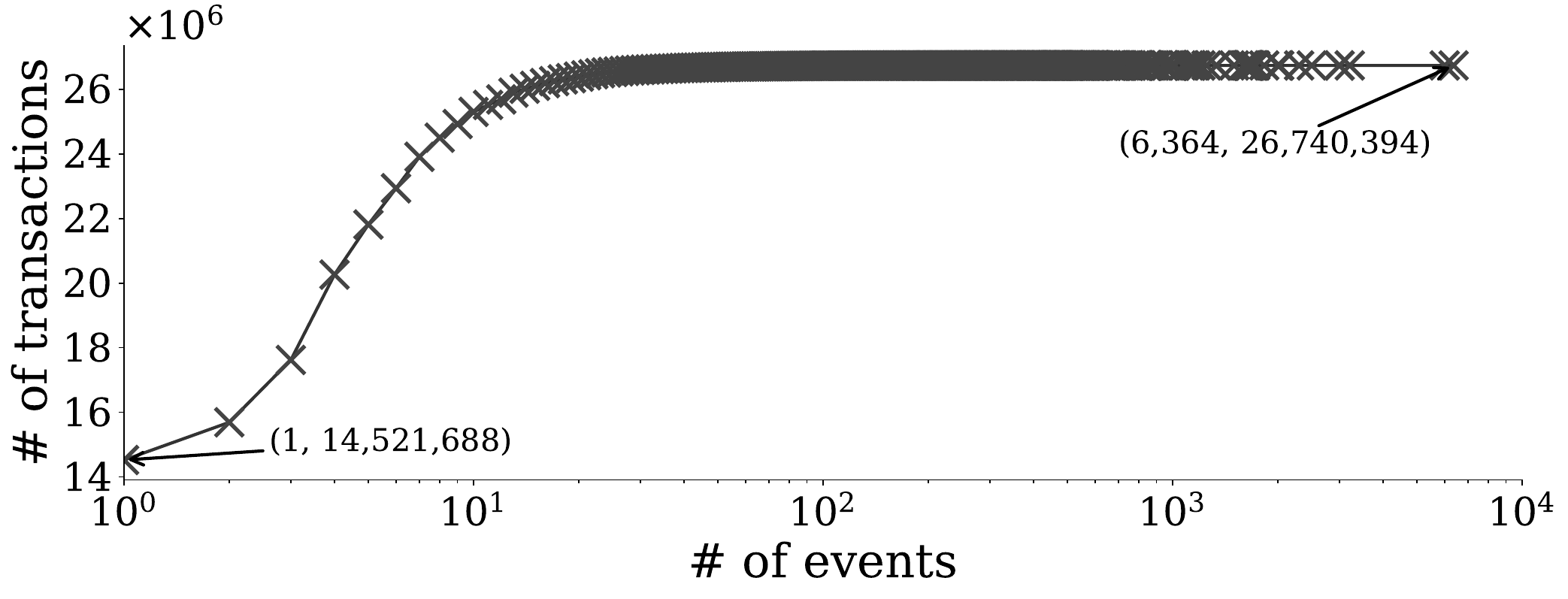}
    \caption{Number of events in each transaction of bundles.}   
    \label{fig_eventcount}
\end{figure}

\noindent
$\bullet$
The number of events in each transaction of bundles. In Fig.~\ref{fig_eventcount}, each cross $(x, y)$ indicates that $y$ transactions contain no more than $x$ events. For 54.31\% %
transactions there is no more than one event in it, and for 31.81\% (8,506,831/26,740,394) transactions there is no event. We find that the transaction\footnote{\href{https://etherscan.io/tx/0xa90088e0848666e93118e2155d336bac0e51afa78373d5e552901d07af1b5911}{0xa90088e0...af1b5911}} emits the most amount of events (i.e., 6,364). 
In the transaction, an NFT contract mints 6,363 NFTs. For each minted NFT, the contract emits a \texttt{\small Transfer} event.

\noindent
\textbf{FP/FN rates.}
We further evaluate the FP/FN rates for the results of new DeFi MEV activity in Table~\ref{tab_kmeans_cnn2d}.
Due to the lack of ground truth for MEV activities, 
we can only evaluate the FP/FN by manually checking each bundle. 
Since to manually analyze all bundles are labor-intensive,
we choose to sample bundles from \mevformula{D$_{\mevsmallformula{Bundle}}$} and manually evaluate the FP/FN results of the sampled bundles.
Specifically,
for each kind of MEV activity, we sample 20 bundles from bundles which contain such kind of MEV activity, 
and we sample 20 bundles from bundles that do not contain the MEV activity.
The results show that there is no FP or FN for results of new DeFi MEV activities in our sampled bundles.

\section{three stealthy attacks in bundles}
\label{sec_twoattacks}

\noindent
-
\textbf{Cream Finance exploitation.} The root cause is that Cream Finance does not set the reentrancy protection and 
\texttt{\small AMP} has a reentrancy vulnerability. 
Specifically, \texttt{\small AMP} contains a hook function \texttt{\small tokensReceived} of ERC777 (\url{https://eips.ethereum.org/EIPS/eip-777}) standard, and the hook function can trigger the execution of the attacker's contract.
The attacker exploited the reentrancy vulnerability in \texttt{\small AMP} token contract, and re-loaned assets from Cream Finance.
Specifically, the adversary first borrowed \texttt{\small AMP} from Cream Finance, and then exploited the reentrancy vulnerability in \texttt{\small AMP} token contract to re-borrow ETH from Cream Finance. Finally, the adversary received profits due to multiple loans with single collateral.

\noindent
-
\textbf{xToken exploitation.}
The root cause is that xToken mints \texttt{\small xSNXa} with relying on the price of \texttt{\small SNX} in Uniswap.
The attacker conducted an indirect price manipulation attack~\cite{wu2021defiranger} on xToken.
Specifically,
the attacker first borrowed \texttt{\small SNX} from flash loans,
and sold the \texttt{\small SNX} to Uniswap aiming at tanking \texttt{\small SNX}'s price.
Then, the attacker used ETH to mint \texttt{\small xSNXa} at the tanked price of \texttt{\small SNX}.
Finally, the attacker burned \texttt{\small xSNXa} and claimed \texttt{\small SNX} tokens. 
The attacker paid back the flash loans and exchanged the remaining \texttt{\small SNX} into ETH as profits.

\noindent
-
\textbf{RigoBlock whitehat rescue.}
Fig.~\ref{fig_rigoblock} shows the code snippets of the RigoBlock.
The comments at Line 1 and 4 show that the two functions, i.e., \texttt{\small setMultipleAllowances} and \texttt{\small operateOnExchange}, 
should be only invoked by \texttt{\small owner}.
But, the developers miss setting the \texttt{\small onlyOwner} modifier~\cite{solidity2022event} for
the first function although \texttt{\small onlyOwner} modifier is set for the second function at Line 5.
The vulnerability allows non-owners to invoke the first function and set allowances for approved tokens to them.
Before the attack transaction, the whitehat invoked the first function and set allowances for them.
Then in the attack transaction, the whitehat drained out six kinds of assets from RigoBlock, and exchanged all of them into ETH.
After the attack transaction, the whitehat communicated with RigoBlock in Twitter and returned the ETH to RigoBlock.

\section{Background on representation learning}
\label{sec_appendix_background}
Representation learning is a class of machine learning methods to automatically discover the features for constructing classifiers or other predictor variables~\cite{bengio2013representation,noroozi2017representation}. Representation learning is fully data-driven and task-oriented, obviating considerable manual efforts for data study and manually extracting features (e.g., feature engineering)~\cite{bengio2013representation,noroozi2017representation}. There are two main types of representation learning, namely supervised representation learning and unsupervised representation learning~\cite{bengio2013representation}. Supervised representation learning learns features from labeled data, such as neural networks, multi-layer perceptrons, and supervised dictionary learning~\cite{bengio2013representation}. Unsupervised representation learning learns features from unlabeled data, such as unsupervised dictionary learning, independent component analysis, automatic coding, matrix factorization~\cite{bengio2013representation}.

\begin{table}[t!]
\centering
\caption{Three DeFi incidents found during the procedure of discovering DeFi MEV activities}
\resizebox{\linewidth}{!}{
\begin{tabular}{|l|c|l|}
\hline
Transaction                                                        & Financial losses (USD) & Description         \\ \hline
\rowcolor{black!20} \href{https://etherscan.io/tx/0xa9a1b8ea288eb9ad315088f17f7c7386b9989c95b4d13c81b69d5ddad7ffe61e}{0xa9a1b8ea...d7ffe61e} & 18.8M & Cream Finance exploitation \\
\rowcolor{black!5} \href{https://etherscan.io/tx/0x7cc7d935d895980cdd905b2a134597fb91004b5d551d6db0fb265e3d9840da22}{0x7cc7d935...9840da22} & 24.5M & xToken exploitation        \\
\rowcolor{black!20} \href{https://etherscan.io/tx/0x5a6c108d5a729be2011cd47590583a04444d4e7c85cd0427071b968edc3bfc1f}{0x5a6c108d...dc3bfc1f} & 42.2K & RigoBlock whitehat rescue  \\ \hline
\end{tabular}
}
\label{table_caseanalysis}
\end{table}

\section{Retraining cost of our model}
\label{sec_appendix_cost}
\noindent
$\bullet$
\textbf{The cost of MEV labeling for bundles}. The MEV activities in bundles can be automatically labeled by summarizing heuristics like existing studies~\cite{qin2021quantifying,ferreira2021frontrunner,wang2021cyclic} to detect known and discovered MEV activities, and hence the MEV labeling is high-efficient. 

\noindent
$\bullet$
\textbf{The computational cost of training the model}. The training of our model only costs a few minutes (< 1 hour) on our server (i.e., Intel Xeon W-1290 CPU with 10 cores at 3.2 GHz).

\begin{figure}[t!]
\small
\begin{lstlisting}[mathescape=true,language=Solidity, frame=none, basicstyle=\linespread{0.6} \fontsize{6}{9}\ttfamily]
/// @dev Allows owner to set allowances to multiple approved tokens with one call.
function setMultipleAllowances(...) {...}
...
/// @dev Allows owner to operate on exchange through extension.
function operateOnExchange(...) onlyOwner {...}
\end{lstlisting}
\captionof{figure}{Code snippets of RigoBlock contract.}
\label{fig_rigoblock} 
\end{figure}

\section{Collecting waiting times}
\label{sec_appendix_waiting_time}

We reused methods in~\cite{liu2022empirical} to obtain a nine-day waiting time dataset for transactions from Mar. 14, 2023, to Mar. 22, 2023.
Specifically, for a transaction \mevformula{TX}, its waiting time = \mevformula{T$_{\mevsmallformula{block}}^{\mevsmallformula{TX}}$} - \mevformula{T$_{\mevsmallformula{mempool}}^{\mevsmallformula{TX}}$},
where 
\mevformula{T$_{\mevsmallformula{block}}^{\mevsmallformula{TX}}$} is the time when \mevformula{TX} is mined,
and \mevformula{T$_{\mevsmallformula{mempool}}^{\mevsmallformula{TX}}$} is the time when \mevformula{TX} first appears in the
mempools of miners/validators.
With the following methods in~\cite{liu2022empirical}, we estimate the earliest time when our nodes monitor a transaction as the \mevformula{T$_{\mevsmallformula{mempool}}^{\mevsmallformula{TX}}$} of this transaction. 
To estimate the time when transactions first appear as precise as possible, we increase the connectivity of our nodes to the Ethereum P2P network by three methods~\cite{liu2022empirical}: i) deploy our nodes to be geographically distributed, ii) configure our nodes with a maximum peer limit of 1,000, and iii) make our nodes to connect to well-known nodes~\cite{liu2022empirical}.
We further estimate the time when the transaction appeared in a block as the \mevformula{T$_{\mevsmallformula{block}}^{\mevsmallformula{TX}}$} of this transaction following methods in~\cite{liu2022empirical}.